\documentclass[twocolumn]{aastex631}

\newcommand{\revise}[1]{\textcolor{black}{#1}}

\usepackage{verbatim}
\usepackage{amsmath}

\begin{document}

\title{The Impact of Fiber Cross Contamination on Radial Velocity Precision}

\correspondingauthor{Chenyang Ji, Sharon X. Wang}
\email{joanneeeji@gmail.com, sharonw@tsinghua.edu.cn}

\author[0009-0005-3557-5183]{Chenyang Ji}
\affiliation{Department of Astronomy, Tsinghua University, Beijing 100084, People's Republic of China}
\author[0000-0002-6937-9034]{Sharon X. Wang}
\affiliation{Department of Astronomy, Tsinghua University, Beijing 100084, People's Republic of China}
\author{Kai Zhang}
\affiliation{Nanjing Institute of Astronomical Optics \& Technology, National Astronomical Observatories of the Chinese Academy of Sciences, Nanjing 210008, People's Republic of China}
\author[0000-0003-3603-1901]{Liang Wang}
\affiliation{Nanjing Institute of Astronomical Optics \& Technology, National Astronomical Observatories of the Chinese Academy of Sciences, Nanjing 210008, People's Republic of China}
\affiliation{CAS Key Laboratory of Astronomical Optics \& Technology, Nanjing Institute of Astronomical Optics \& Technology, Nanjing 210042, China}
\affiliation{University of Chinese Academy of Sciences, Beijing 100049, China}



\begin{abstract}

High-resolution spectrographs with precise radial velocity (PRV) capabilities require careful considerations in instrumental design and data processing in order to reach the 10~cm/s-level precision, which is needed for detecting Earth-like planets. 
In this work, we investigate the impact of fiber cross contamination on the RV precision via simulations, as modern PRV spectrographs often have multiple fiber traces on their spectral images. We simulated extracted 1-D spectra under the preliminary design of CHORUS, short for the Canary Hybrid Optical high-Resolution Ultra-stable Spectrograph, a dual-arm PRV spectrograph under construction for the Gran Telescopio de Canarias. 
We considered two types of fiber cross contaminations: contamination from calibration traces to neighboring science traces (or cal-sci contamination) and between science traces (or sci-sci contamination). We present results in four different scenarios: photon noise only, cal-sci contamination only, sci-sci contamination only, and all effects combined. 
For the preliminary design of CHORUS, we estimated that the cal-sci contamination fraction is smaller than $0.0001\%$ in flux across the whole CCD for either arm, resulting in a negligible impact on the RV precision. Assuming worst-case scenarios, we estimated the sci-sci contamination to be up to $0.1\%$ in some traces, corresponding to an additional RV error of up to $10$~cm/s. 
We demonstrate the importance of considering fiber-trace spacing and cross contamination in PRV spectrographs, and we recommend careful design, operation, and spectral extraction algorithms to minimize and mitigate cross contamination to achieve the best possible instrumental RV precision.

\end{abstract}



\section{Introduction \label{sec:intro}}

The first detection of an exoplanet around a Sun-like star was made with radial velocity (RV) precision of tens of meters per second \citep{1995Natur.378..355M}. Ever since this landmark discovery, exoplanets of various masses and orbital periods have been identified, and we have entered the era where finding Earth-like planets in the Habitable Zone of nearby Sun-like stars is almost within reach and is among one of the important science questions within this decade \citep[e.g.,][]{2021pdaa.book.....N}. The RV method remains an important venue for discovering Earth-like planets, and the RV precision has been steadily improved by instrumental advancements such as stabilized spectrographs and accurate wavelength calibrations \citep[e.g.,][]{2016PASP..128f6001F, 2017RNAAS...1...51W}, as well as more advanced data analysis and statistical methods \citep[e.g.,][]{2021A&A...653A..43C, 2022AJ....164...84A, 2023AnRSA..10..623H}. 

To detect Earth-like planets in habitable zone \citep{1993Icar..101..108K, 2013ApJ...765..131K}, long-term precise radial velocity (PRV) must reach a precision of $< 10 \mathrm{cm/s}$, which poses challenges to instrumentation, data analyses, as well as understandings of stellar astrophysics \citep[e.g., ][]{2016PASP..128f6001F}. Exoplanet searches using PRV spectrographs would have to consider a multitude of RV error sources stemming from the instrument \citep[e.g.,][]{2016SPIE.9908E..6PH}, fundamental limits set by the stars \citep[e.g.,][]{2015PASP..127.1240B, 2020ApJS..247...11R, 2019A&A...632A..37B, 2020AJ....159..235L}, and also intricate effects such as contamination light from the scattered Sunlight \citep{2020AJ....159..161R} or telluric lines \citep[e.g.][]{2022AJ....164..211W, 2022AJ....164..212L,2022A&A...666A.196A}.

In this paper, we investigate one of these intricate effects in PRV spectrographs -- the effect of cross contamination between fibers or fiber traces. Most PRV spectrographs on eight-meter class telescopes employ fibers: HPF on HET \citep{2014SPIE.9147E..1GM}, MAROON-X on Gemini North \citep{2018SPIE10702E..6DS}, ESPRESSO on VLT \citep{2013Msngr.153....6P, 2014AN....335....8P, 2021A&A...645A..96P}, KPF on Keck \citep{2016SPIE.9908E..70G}, and iLocater on LBT \citep{2016SPIE.9908E..19C}, to name a few. They either use single-mode fibers (iLocator), which benefit in producing collimated light but with low coupling efficiency \citep{2015ApJ...814L..22H, 2016SPIE.9908E..19C}, or multi-mode fibers combined with image or pupil slicers or an additional slit to enhance spectral resolution. The choice of image (KPF) or pupil slicer (ESPRESSO and MAROON-X) would lead to a group of multiple fiber traces for each order, which would take up more real estate on the detector. In cases where the traces are close to each other in pixel space, cross contamination might be an issue.

We study the effect of fiber cross contamination under the context of the preliminary design of a new upcoming PRV spectrograph: CHORUS, Canary Hybrid Optical high-Resolution Ultra-stable Spectrograph. CHORUS is a high-resolution spectrograph currently under construction by the National Observatories of the Chinese Academy of Sciences (NAOC-NIAOT), as part of a collaboration with the Gran Telescopio Canarias (GTC) established by a 2016 agreement \footnote{For more information, see the CHORUS homepage: \url{https://www.nao.cas.cn/gtc/}}. The preliminary optical design of CHORUS includes a pupil slicer, which divides the light beam in the visible band (430 -- 780\,nm) into three traces, resulting in the CCD layout shown in Figure~\ref{fig:ccd}. This three-slice design offers a higher spectral resolution of around 120,000 than a two-slice design (with R$\sim$90,000), but at the cost of narrower separations between the traces that could lead to potential issues of cross contamination. It is thus important to assess the potential impact of fiber cross contamination in order to achieve the instrumental RV precision goal of $10\mathrm{cm/s}$, which is the focus of this paper.

The paper is organized as follows: Section~\ref{sec:meth} describes how we carried out the simulation, including spectrum simulation, cross contamination injection, and RV extraction; Section~\ref{sec:res} reports our results under different contamination scenarios; Section~\ref{sec:dc} includes discussion, future work, and a final conclusion.

\section{Methodology \label{sec:meth}}

Overall, our simulations assumed a specific spectral format (i.e., orders, wavelength range, resolution, etc.), and in this paper, we adopted the format from the preliminary optical design of CHORUS as a representative setup for a fiber-fed PRV spectrograph. We synthesized multiple 1-D extracted spectra for a few main-sequence stars with different effective temperatures. We tested the effects on the RV precision of two types of cross contaminations: one from the adjacent calibration line to the synthetic spectra, and one between the synthetic spectra within one order. We used the forward modeling method to extract the RV of each spectral order in each simulated observed spectrum by comparing the template and synthetic spectra. RV precision is estimated by taking the standard deviation of RVs throughout all the observations. We describe the details of these steps in the following subsections.

\subsection{Input Spectral Format and Settings \label{subsec:input}}
We assume a dual-arm fiber-fed high-resolution spectrograph with the red arm covering 5370 to 7804\AA~and the blue arm covering 4088 to 5444\AA. We adopted a uniform spectral resolution of R=120,000 across the entire wavelength range of the spectrograph, although in reality, spectral resolution \revise{increases from blue to red} within an order and also across different orders (by $\sim$10--20\%). As RV precision depends on spectral resolution, this would translate into a change in the intrinsic photon-limited RV precision versus wavelength. However, since we estimated the RV precision using information combined from all orders, this would be equivalent to setting the resolution to a constant value. We thus neglected the change of spectral resolution versus wavelength for simplicity in our simulations.

\begin{figure*}[ht]
    \centering
    \includegraphics[width=1\linewidth]{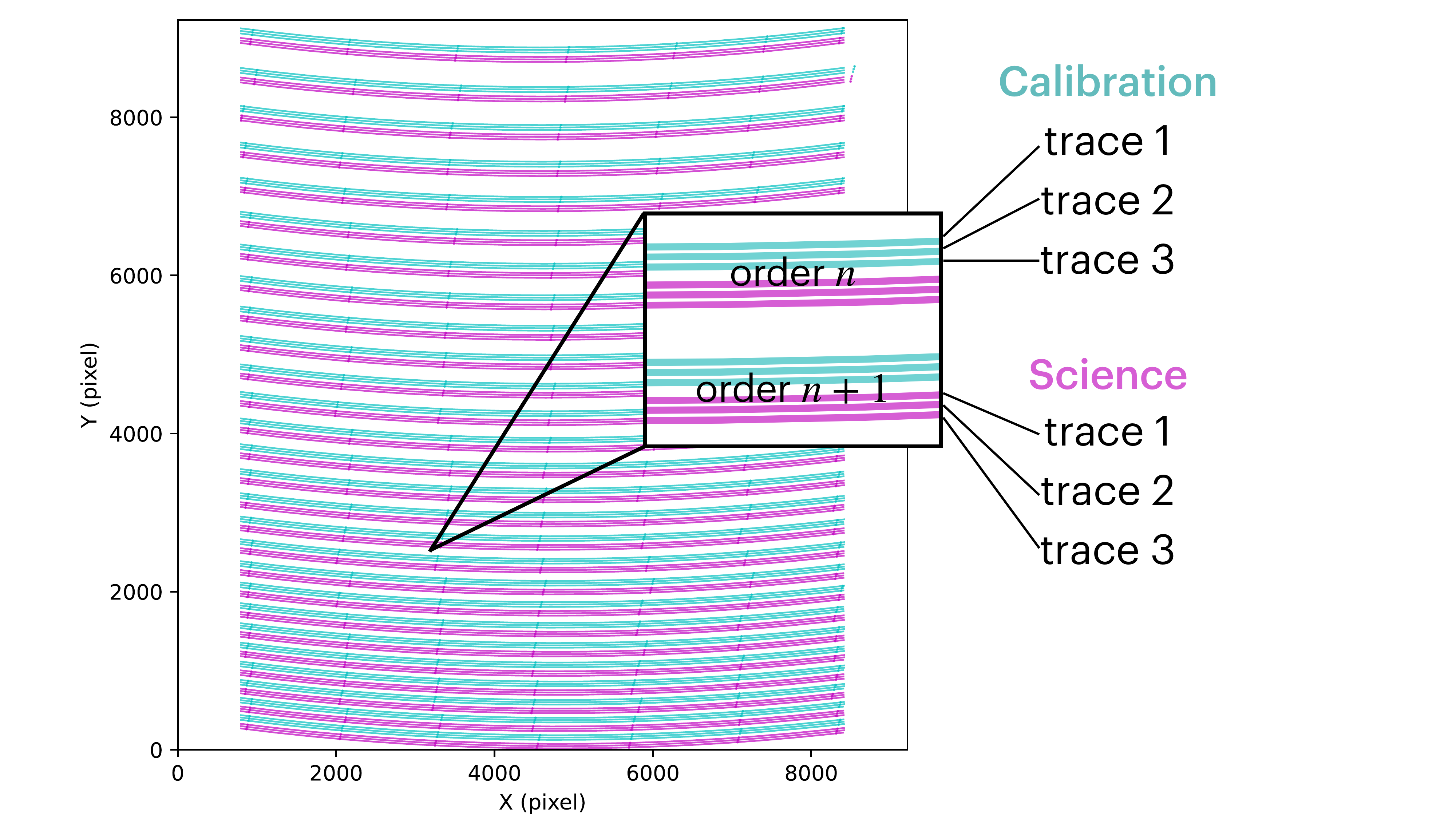}
    \caption{Illustration of the red-band CCD layout of CHORUS according to its preliminary design. There are 27 orders in the red band, each containing three calibration traces and three science traces. The blue-band CCD has a similar layout but with 28 orders.}
    \label{fig:ccd}
\end{figure*}

As commonly seen in high-resolution, fiber-fed spectrographs behind large telescopes (e.g., VLT ESPRESSO, Gemini MAROON-X, or Keck KPF), there are multiple slices for each spectral order, either as a result of a pupil slicer or an image slicer. We assume a three-slice format at the pupil, where the input stellar light is basically evenly split into three portions before going into the spectrograph. This results in three traces in each spectral order of the science spectrum, as well as in each of the calibration spectral orders. A layout of the spectral format is in the left panel of Figure~\ref{fig:ccd}, with the red-band CCD as an example. Each order on CCD contains three calibration and three scientific traces. 

\begin{figure*}[ht]
    \centering
    \includegraphics[width=1\linewidth]{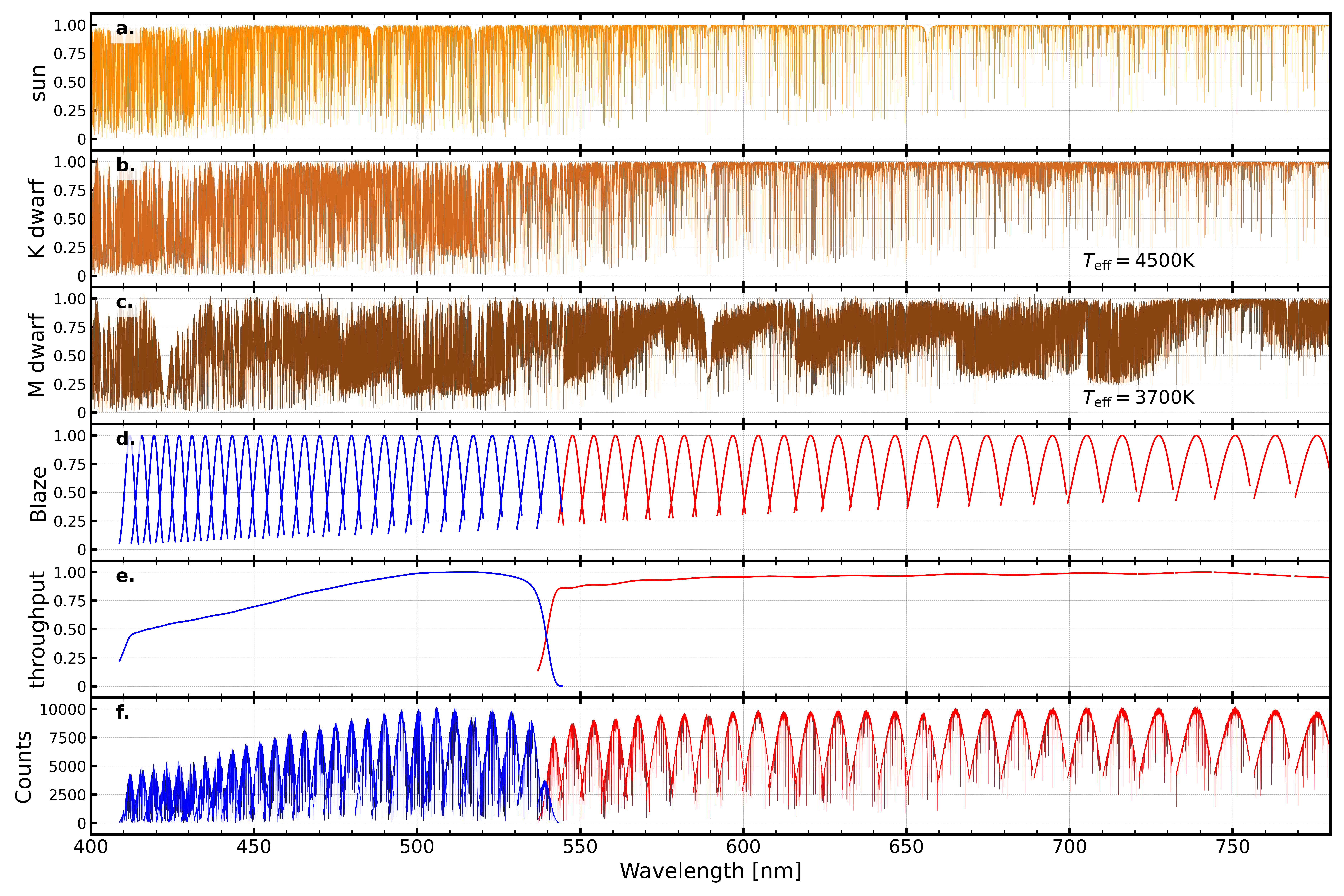}
    \caption{Illustration of our input stellar spectra (panel a--c), the normalized blaze functions and the throughput curves (panel d and e), and an example of the final simulated 1-D solar spectrum with peak SNR=100 (f). }
    \label{fig:intro}
\end{figure*}

For each order, the wavelength range, Blaze function, and optical throughput profile are determined by the optical settings of the echelle. The blaze function and the throughput profile (normalized to unity at maximum) are presented in panels d and e in Figure~\ref{fig:intro}. We assume an echelle grating with a groove density of 31.6\,l mm$^{-1}$, blaze angle of 75$^\circ$, off-Littrow angle ($\theta$) of 0\,$^\circ$, and out-of-plane angle ($\gamma$) of 1.0$^\circ$. The spectral format was then calculated in \texttt{Zemax}\footnote{\url{https://www.ansys.com/products/optics/ansys-zemax-opticstudio}}, with each order having seven ray traces providing the wavelength solution of 7 pixels for each order trace. We then computed the wavelength of each pixel by interpolating and extrapolating these 7 points using a second-order polynomial.

Each order (or each trace of order) has an effective width of 8000 pixels, which is shorter than the commonly adopted 9k by 9k CCD for such spectrographs, as we assumed the detector is underfilled and the edges are neglected given the low signal-to-noise ratio (SNR; see the left panel in Figure~\ref{fig:ccd}). We assume a read-out noise of $\sigma_e = $ 2.5 electrons per pixel (typical for a relatively slow read-out mode). 

\subsection{Simulating Spectra} \label{subsec:simulate}

With the spectral format described above, we then simulated the 1-D extracted spectrum for each trace within each order. As we did not simulate the 2-D spectral images, the spacing between each of these order traces did not enter our synthetic spectra until we calculated and added the contaminated light from neighboring slices, which we describe later in Section \ref{subsec:cont}. 

We chose to simulate three stellar spectral types in the main sequence, including a solar twin, a K dwarf with $T_\mathrm{eff}=4500\mathrm{K}$, and an M dwarf with $T_\mathrm{eff}=3700\mathrm{K}$, representing typical targets observed by such spectrographs. We adopted the computed solar spectrum with a resolution of 500,000 from the Kurucz model\footnote{Downloaded here in 2022: \url{http://kurucz.harvard.edu/stars/sun/}} \citep{2005MSAIS...8...14K}. We set $\log g=4.5$ and $\mathrm{[Fe/H]}=0 \, \mathrm{dex}$ for both K and M dwarfs and obtained their spectra from the PHEONIX model library\footnote{Downloaded here in 2023: \url{https://www.astro.uni-jena.de/Users/theory/for2285-phoenix/grid.php}} \citep{2013A&A...553A...6H}, also with a resolution of 500,000. The Kurucz's solar spectrum provided a continuum-normalized version, and we normalized the K and M dwarfs' PHEONIX model spectra using \texttt{iSpec}\footnote{\url{https://github.com/marblestation/iSpec}} \citep{2014A&A...569A.111B, 2019MNRAS.486.2075B}. We added no additional line broadening due to stellar rotation, i.e., assuming $v \sin i = 0$~km/s. We then up-sampled the model spectra by a factor of 5 using cubic interpolation to improve numerical stability for our simulations. Illustrations for the three templates are in panels a to c in Figure~\ref{fig:intro}. We used these up-sampled normalized high-resolution spectra, which have wavelengths set in the stellar rest frame (i.e., $z=0 \mathrm{km/s}$), as our input spectra for generating the simulated observed spectra. We refer to these model spectra as the ``input spectra" or ``template spectra" below. 

Our overall spectral simulation process goes as follows: we started with a minimum unit of a spectral trace, looping through each of the three traces in each order and continuing to other traces in all orders in the red and blue bands, respectively. 
The three traces within an order are identical in terms of the incoming stellar light into the spectrograph, but they differ in their exact wavelength solution on the CCD grid due to their relative displacement on the CCD, as well as the exact added photon or readout noises and contamination (but with the same SNR; more later in this section and Section~\ref{subsec:cont}). For each spectral trace within an order, we first took a section from the input spectrum according to the wavelength range of this order, added an extra 2\AA~on both ends, and then convolved the spectrum with a line-spread function. We then re-sampled the convolved spectrum onto the 1-D CCD grid for this trace (8000 pixels wide) and added photon and readout noise. We describe the simulation process in more detail below.

We convolved the template spectrum with a spectral point-spread function (PSF; also called the line-spread function or the spectrograph response function). The exact format of the spectral PSF does not matter since we assume a non-changing PSF and also perfect knowledge of the PSF in our simulation to avoid additional RV errors. We assumed a Gaussian function for PSF, with its full width at half maximum (FWHM) set by the spectral resolution (R=120,000) and the median wavelength of each order, $\lambda_m$. 
Mathematically,
\begin{equation}
    F_\mathrm{120k} = F_\mathrm{temp} \otimes \mathrm{PSF},\ {\rm where~PSF} = \frac{\exp \left(-\frac{1}{2} \frac{x^2}{\sigma^2}\right)}{\int \exp \left(-\frac{1}{2} \frac{x^2}{\sigma^2}\right) \, {\rm d}x} \, ,\label{eqn:psf}
\end{equation}
\begin{equation}
    \text{with }\, \sigma = \frac{\mathrm{FWHM}}{2.35} = \left( \frac{\lambda_m}{R} \large/\Delta \lambda \right) \large/ 2.35. \label{eqn:sigma}
\end{equation}
Here, $F_\mathrm{temp}$ is the input template spectrum for this order, and $\Delta \lambda$ refers to the median sampling rate of the template spectrum of this order, estimated using the median of the wavelength differences between two adjacent points in the template spectrum $F_\mathrm{temp}$. $\Delta \lambda$ here converts the unit of $\sigma$ from wavelength (\AA, as set by $\lambda_m/R$/2.35) to the numerical grid of the spectral template. $R$ is the spectral resolution. The factor 2.35 is the analytical factor converting the FWHM of a Gaussian to its standard deviation $\sigma$ ($\sqrt{8\ln{2}} \simeq 2.35)$. We constructed the PSF on a wide grid with a half-width well beyond the 3-$\sigma$ range (i.e., range of $x$ in Eq.~\ref{eqn:psf} $\gg 6\sigma$).

We then resampled the convolved spectrum onto the 1-D CCD pixel grid according to the wavelength solution of this trace. Next, we added photon counts to the resampled spectrum to synthesize the final observed spectrum, taking into account instrument throughput curves, blaze functions, photon noise, and readout noise. As a result, we have
\begin{equation}
    \begin{aligned}
        F_\mathrm{pho}(x) \sim & \mathrm{Poisson}\left[F_\mathrm{120k}(x) \cdot \mathrm{SNR}^2 \cdot B(x) \cdot f_\mathrm{tp}(x) \right] \\
        & + 17 \cdot \mathcal{N}(\mu=0, \sigma^2=2.5^2). 
        \label{eqn: resample}
    \end{aligned}
\end{equation}
Here, $x$ is the pixel range, SNR stands for the peak signal-to-noise ratio over the whole spectral range, $B(x)$ is the normalized blaze function for this order, and $f_\mathrm{tp}(x)$ is the relative instrument throughput for this order, which is normalized to 1.0 at peak over the whole spectral range. 
Finally, the total noise added for each pixel in this trace is a combination of Poisson noise and a normally distributed readout noise added up for all extracted pixels along the cross-dispersion direction (assumed to be 17 pixels from the preliminary design of CHORUS). This gives the final simulated spectrum for this trace. An example of the simulated 1-D spectrum with a peak SNR of 100 is shown in panel f in Figure~\ref{fig:intro}.

We then iterate through each trace of each order and for all orders in the red and blue bands. We typically generate 100 simulated spectra for one set of simulations and use the RV scatter among these simulated observations to assess the RV precision. \revise{We set a constant peak SNR for one setting in our simulation (introduced later), based on the assumption that photon counts of the three traces within one order are the same. However, the photon counts and thus SNR would normally differ between traces within an order, depending on how the slicing is performed. In the case of the preliminary design of CHORUS, the sliced pupil is an elliptical aperture with a major-to-minor axis ratio of $4.5:1$, and it is trisected along its major axis. Consequently, the three traces (or photon counts) are not strictly equal based on area. We will discuss the caveats of this assumption in Section~\ref{subsec:cf}.} 

\subsection{Cross Contamination} \label{subsec:cont}

Contamination between the fiber traces inherently arises due to the probability distribution of photons along the cross-order direction, which we will refer to as ``cross contamination" below. This issue could become significant when the separations between traces are small, causing a noticeable leakage of photons from neighboring traces onto each other. In this section, we describe our assumptions and simulation processes in order to quantify the influence of cross contamination on RV precision.

We made several assumptions regarding how cross contamination happens and the spectral extraction processes to simplify our simulations. First of all, since the separation between traces and orders varies across different positions on the CCD (see the left panel in Figure~\ref{fig:ccd}), the exact amount of contamination also varies across different traces/orders or pixels within each trace. The most accurate estimate of the cross contamination between different traces/orders would require a full 2-D simulation of the spectrum on the CCD and an emulation of the spectral extraction process. Instead of pursuing this full set of simulations, we adopted a simple assumption that all neighboring traces have a fixed amount of cross contamination onto each other (in terms of a fraction of the photon counts, referred to as the ``contamination fraction" hereafter), and then we vary the amount of cross contamination to examine how the RV precision would change. 

Moreover, we assume the extraction windows for each pair of neighboring traces were divided at the median of the clear pixels between traces. In reality, the extraction window is almost certainly smaller than assumed here. We also assume that no spectral rectification was done to correct the tilt between the cross-dispersion direction and the pixel axis and that the extraction was done by simply adding the photons along the pixels. Therefore, we are estimating the ``maximum amount of contamination possible", in some sense. Finally, in the equations and figures below, we adopted the CCD 2-D coordinates $(i,j)$, with $i$ as the x-pixel location (roughly in the dispersion direction along the CCD rows; see Figure~\ref{fig:ccd}) and $j$ as the y-pixel location (roughly in the cross-dispersion direction along the CCD columns).

\begin{figure}[ht]
    \centering
    \includegraphics[width=1\linewidth]{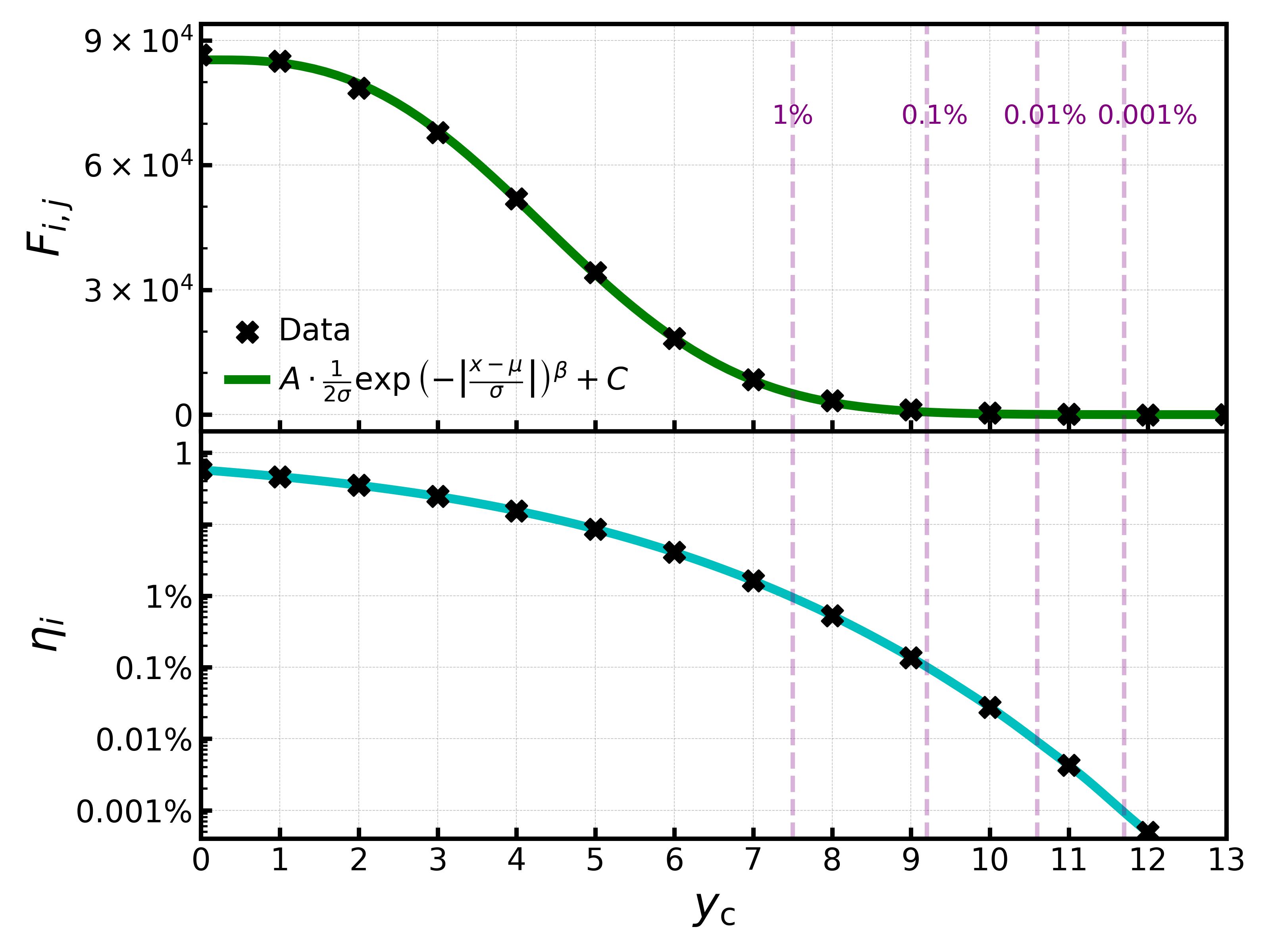}
    \caption{The original photon counts at location $(i,j)$ $F_{i,j}$ and fraction of contaminating photons $\eta_i$ along the cut-off pixels $y_c$ from the profile center in the cross-disperser direction. Data was downloaded from ESO archive observed by ESPRESSO. \textbf{Upper panel:} The black crosses show the column-stacked photon counts of one trace from one archival observation by ESPRESSO. The green line is the best generalized-Gaussian fitting for these counts, following the expression in legend with $\beta \sim 2.66$. \textbf{Lower panel:} The black crosses show the fraction of counts integrated from the cut-off pixel to the rightmost of this trace. The cyan line is the interpolation of these points. Contamination fractions of 1\%, 0.1\%, 0.01\%, and 0.001\% are marked with purple dashed lines in both subplots. See Section \ref{subsec:cont} for details.}
    \label{fig:cont_frac}
\end{figure}

The first critical element in estimating the amount of cross contamination is the spectral PSF along the cross-dispersion direction. We illustrate the functional form of the spectral PSF we adopted based on fitting the ESPRESSO on-sky data in Figure~\ref{fig:cont_frac}. We selected an archival raw spectrum obtained by ESPRESSO on HIP~41378 on December 1st, 2022\footnote{Downloaded in 2024 from \url{https://archive.eso.org/eso/eso_archive_main.html}, program ID 105.20G0.001. Observed with the HR mode with a single telescope (``SINGLEHR").}, sliced the image along 30 columns of pixels near the center of one arbitrary bright trace, stacked the counts from these columns (denoted as effectively the ``$i$-th column below''), and plotted the distribution of photon counts in the y-pixel direction in the upper subplot of Figure~\ref{fig:cont_frac}, starting from the center of the trace towards the edge (since the shape is symmetric). We fitted the photon distribution using three different profiles: a Gaussian distribution, a Gaussian convolved with a top-hat function, and a generalized Gaussian profile (equation in Figure~\ref{fig:cont_frac} legend). Among these, the generalized Gaussian profile provided the best fit, capturing the flatter peak of the distribution. The resulting fit parameters were $\sigma \sim 4.96$, $\mu \sim 75.18$, $A \sim 840000$, $C \sim 18000$, and $\beta \sim 2.66$, with the best-fit plotted in a green thick line. We then adopted this generalized Gaussian profile as the cross-dispersion PSF in our simulation.

Next, we can link the contamination fraction to the separation between two neighboring traces in pixels, assuming a simple spectral extraction procedure where photon counts are summed along the $j$-direction (the cross-dispersion direction) within the extraction window. The contamination fraction at the $i$-th column between traces, denoted as $\eta_i$, is then estimated based on the fiber trace separation and the spectral PSF in the cross-dispersion direction. We defined the ``cut-off pixel" $y_\mathrm{c}$ as the edge of the spectral extraction window, meaning the furthest pixel from the spectral trace center that was included in the extraction. The cross contamination fraction from the current trace onto the neighboring trace is then simply the portion of photon counts beyond $y_\mathrm{c}$ in the current trace. Mathematically, the contamination fraction can be estimated by:
\begin{equation}
    \eta_i (y_\mathrm{c}) = 
    \displaystyle \sum_{j=y_\mathrm{c}}^{j=\infty} F_{i,j}\Big/
         \displaystyle \sum_{j=-\infty}^{j=\infty} F_{i,j}
    \label{eqn:frac}
\end{equation}
In this equation, $F_{i,j}$ represents the photon count at the pixel coordinate ($i$, $j$) on the CCD. We integrate to infinity for simplicity, although in reality, one should integrate within a finite window. In practice, $\eta_i$ varies across different $i$ positions within the trace and between traces or orders. However, for simplicity, we assume a constant value of $\eta$ within any simulated spectrum, and we vary $\eta$ to examine how the RV precision changes versus various fiber spacings, as mentioned earlier.

\begin{figure}[ht]
    \centering
    \includegraphics[width=1\linewidth]{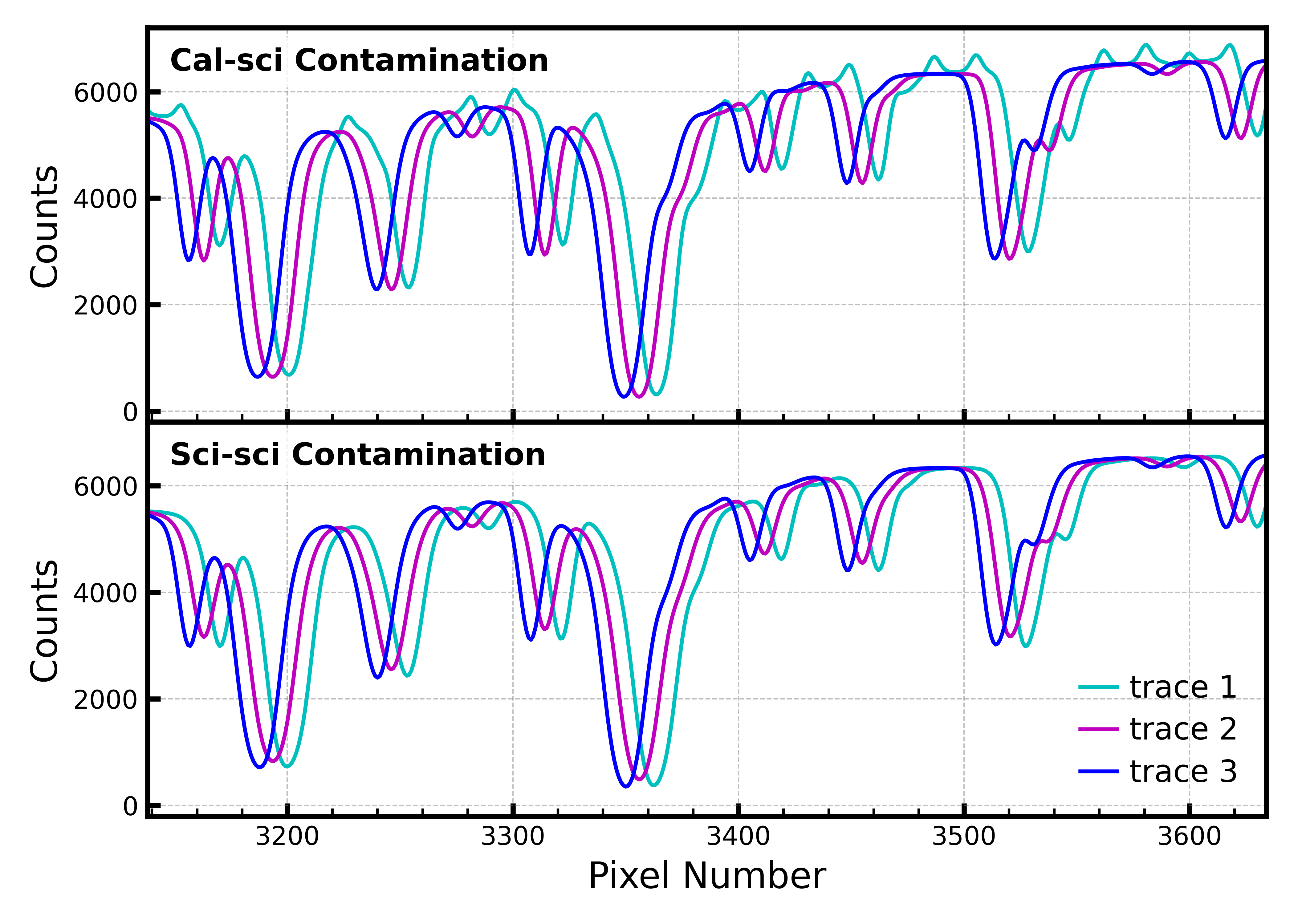}
    \caption{An illustration of how cal-sci (top panel) and sci-sci (bottom panel) contaminations affect the extracted science spectrum (solar template as an example). Each panel shows the extracted spectrum in each trace of an arbitrary order in terms of the total photon counts versus pixel number along the dispersion direction on the CCD. Both spectra were simulated with a spectral resolution of 120,000 and peak photon counts of 10,000, without adding any photon noise. For illustrative purposes, we assumed an unrealistically large contamination fraction of 10\% in both panels.}
    \label{fig:contamination_pixel}
\end{figure}

The lower panel of Figure~\ref{fig:cont_frac} illustrates how the contamination fraction $\eta_i$ changes with the cut-off pixel $y_\mathrm{c}$, with the values between the integer pixel points interpolated using a cubic spline (solid cyan line). For example, when the spectral extraction window of the neighboring trace starts at $y_\mathrm{c}=5$, a contamination fraction of about 10\% is introduced. This value is relatively large and unrealistic as the extraction window for the neighboring trace would not be set so close to the center current trace, but we use this example to illustrate the spectra with $\eta = 10\%$ contamination in Figure~\ref{fig:contamination_pixel}. In this study, we explore how the RV precision would change with five different contamination fractions: 0.001\%, 0.01\%, 0.1\%, 1\%, and 10\%, corresponding to a cut-off pixel at roughly 12, 10.7, 9.2, 7.5, and 5.0 from the center of the neighboring trace, respectively (purple dashed vertical lines in \revise{Figure~\ref{fig:cont_frac}}). Note that the pixels quoted here correspond to the cross-dispersion PSF of this particular piece of an ESPRESSO spectrum, and the pixel values are roughly larger by a factor of 1.5 for the optical setup of CHORUS (due to a difference in the dimension of the echelle grating).

We considered two types of cross contamination in our simulation: one is the contamination from the calibration trace to its closest neighboring science trace (referred to as cal-sci contamination), and the other is the contamination between the three science traces (referred to as sci-sci contamination). Given that the spacings between different orders are relatively wide (Figure~\ref{fig:ccd}), we only consider cal-sci contamination between the closest pairs of cal-sci traces and the sci-sci contamination within each order.

For cal-sci contamination, we generated a set of calibration spectra assuming a Fabry-P\'erot (FP) interferometer providing the wavelength calibrations. The theoretical expression for the normalized flux of an FP is given by:
\begin{equation}
F_\mathrm{FP} = \frac{1}{1+f \sin^2 \left(\frac{2\pi L}{\lambda}\right)}, \quad \text{with} \, f=\frac{4r}{(1-r)^2}, \label{eqn:fp}
\end{equation}
where $r$ is the reflectivity of the cavity flats, $L$ is the cavity length, and $f$ is the Lorentzian finesse. We used $L = 7.8 \mathrm{mm}$ and $f = 18$ from the preliminary design of the CHORUS FP. The SNR of the FP spectrum was set to be the same at peak as the science spectrum for simplicity, since in real observations, the brightness of the calibration lamp can be adjusted to match the science spectrum. The simulated calibration spectra $F_\mathrm{cal}$ can then be broadened according to Eqn.~\ref{eqn:psf} and incorporate photon noise according to Eqn.~\ref{eqn: resample}. Then we multiply $F_\mathrm{cal}$ with a cal-sci contamination fraction $\eta_{\text{cs}}$ and arrive at the final extracted, contaminated spectrum as:
\begin{equation}
    F_\mathrm{cont} (x) = F_\mathrm{pho}(x) + \eta_\mathrm{cs} \cdot F_\mathrm{cal}(x)
    \label{eqn:cal}
\end{equation}
The cal-sci contamination to science traces 2 and 3 is negligible compared to science trace 1, which is closest to the calibration traces and therefore most affected (see the zoom-in panel of Figure~\ref{fig:ccd}). The upper panel in Figure~\ref{fig:contamination_pixel} shows the extracted spectrum for each trace of an arbitrary order. We simulated the science spectra with a spectral resolution of 120,000, setting the peak photon counts to 10,000, without introducing photon noise, in order to isolate the effect of contamination. For illustrative purposes, we used an unrealistically large cal-sci contamination fraction of 10\%. As a result, line deformation due to cal-sci cross contamination is quite visible in the extracted spectrum of science trace 1 (cyan line).

For sci-sci contamination, the same y-pixel column in different traces within a single order could correspond to different wavelengths due to misalignments between the cross-dispersion direction and the y-axis. Therefore, sci-sci contamination would exist in all science traces (under our assumption of simple extraction along the y-pixels). The lower panel of Figure~\ref{fig:contamination_pixel} shows the effect of adding 10\% sci-sci contamination (with all other settings the same as in the upper panel). In this case, all three traces exhibit distortions in their line shapes, manifested as variations in absorption depths that should otherwise be identical.

In conclusion, science trace 1 is contaminated by calibration trace 3 and science trace 2; science trace 2 is contaminated by science trace 1 and 3; science trace 3 is contaminated by science trace 2. The expressions for these contaminated spectra are:
\begin{equation}
    \begin{cases}
        F_\mathrm{sci,1} = F_\mathrm{pho,1} + \eta_\mathrm{cs} \cdot F_\mathrm{cal,3} + \eta_\mathrm{ss} \cdot F_\mathrm{pho,2} \\
        F_\mathrm{sci,2} = F_\mathrm{pho,2} + \eta_\mathrm{ss} \cdot F_\mathrm{pho,1} + \eta_\mathrm{ss} \cdot F_\mathrm{pho,3} \\
        F_\mathrm{sci,3} = F_\mathrm{pho,3} + \eta_\mathrm{ss} \cdot F_\mathrm{pho,2}
    \end{cases}
\end{equation}
Here, $F_\mathrm{cal,3}$, $F_\mathrm{pho,1}$, $F_\mathrm{pho,2}$, and $F_\mathrm{pho,3}$ are the input cal or sci spectra without any contamination. $\eta_\mathrm{cs}$ and $\eta_\mathrm{ss}$ stand for cal-sci and sci-sci contamination fractions and have a fixed amount throughout the whole CCD. The final spectrum of each trace would be a joint version of two panels in Figure~\ref{fig:contamination_pixel}.





\subsection{Extracting RVs and Assessing Precision} \label{subsec:RVextra}

The RV in each trace was derived by using the least-squares method to forward model the data set (our simulated spectrum) with the template spectrum. To prepare the simulated spectrum for RV extraction, we first normalized the spectrum of each trace, $F_{\rm pho}$ (Eqn.~\ref{eqn: resample}), by dividing the blaze and throughput profile added before, which means no additional error was induced in continuum normalization. 

To create a model of each spectral trace, we started with the input template spectrum with the matching wavelength range plus an additional 2\AA~segment on each end to avoid edge effects.  We then convolved this piece of input spectrum with the same PSF we used in Section~\ref{subsec:simulate} and Eq.~\ref{eqn:psf} to reach a resolution of 120,000 and arrived at the model spectrum $F_\mathrm{120k}$. Then we Doppler-shifted the input spectrum and evaluated the $\chi^2$ of this model:
\begin{equation}
    \chi^2(z) = \sum_x \left[W\cdot\left(F_\mathrm{120k}(\lambda_z) - F_\mathrm{norm}\right)^2\right],\ {\rm where~} \lambda_z = \frac{\lambda}{1+z/c}, \label{eqn:LMS}
\end{equation}
\begin{equation}
    \text{with }\, W = \frac{1}{\sigma_\mathrm{norm}^2} = \left(\frac{f_\mathrm{tp} \cdot B \cdot \mathrm{SNR}^2}{\sqrt{F_\mathrm{pho}}}\right)^2. \label{eqn:weight}
\end{equation}
Here, $z$ is the Doppler shift (or RV) to fit, $c$ is the speed of light, and $\sigma_\mathrm{norm}$ is the uncertainty in the normalized spectrum. $W$ is the statistical weight of each point (being the inverse of the variance). $\sigma_\mathrm{norm}$ is defined by the normalized photon noise, which is the Gaussian error ($\sqrt{F_\mathrm{pho}}$) divided by the normalizing profile ($f_\mathrm{tp} \cdot B \cdot \mathrm{SNR}^2$; see Eq.~\ref{eqn: resample}).
The best-fit Doppler shift or RV of this trace was estimated by minimizing $\chi^2(z)$. We assume perfect knowledge of the PSF and wavelength solution to avoid additional RV errors. We have validated that our pipeline returns the perfect RV solutions when photon noise is not added. 

\begin{figure*}[ht]
    \centering
    \includegraphics[width=0.6\linewidth]{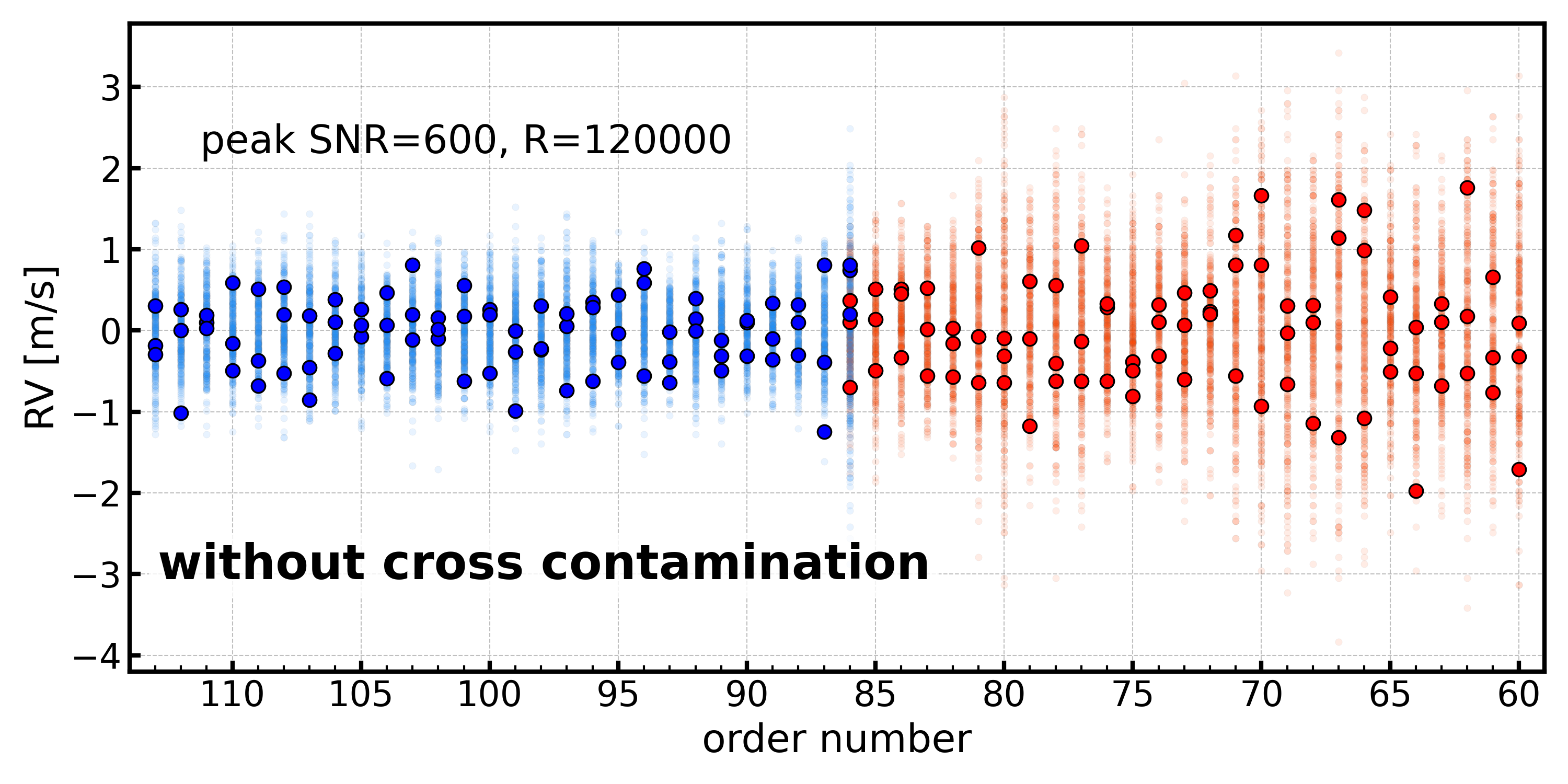}
    \includegraphics[width=0.38\linewidth]{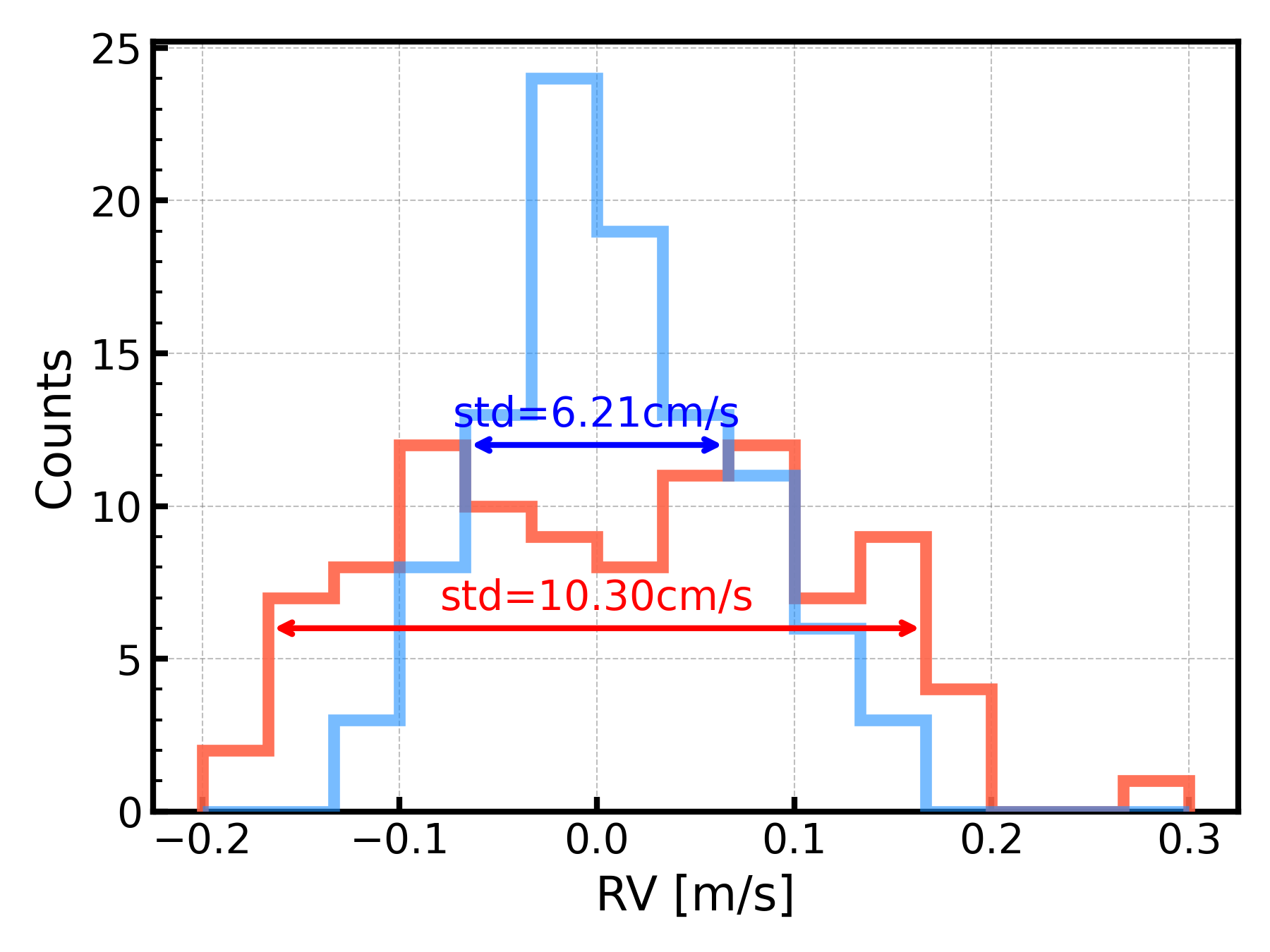}
    \caption{\textbf{Left:} Scatter plot showing the RVs derived from each of the three traces in each spectral order throughout the 100 simulated observations. The dark blue and red filled circles are the RVs of the first observation in the blue and red bands, respectively, and the light points in the background are the RVs for the rest of the observations. The final RV of the blue or red band in each observation was estimated using the median of the RVs in all traces (e.g., median of all the dark blue or red points for observation \#1). \textbf{Right:} Histogram of the final RVs derived from the 100 simulated observations in the blue and red bands, respectively. We took the standard deviation of these 100 derived RVs to represent the RV precision of the blue/red band. Both panels used a series of simulated observations for the sun with a spectral resolution of 120,000 and a peak SNR of 600 without contamination added. See Section~\ref{subsec:RVextra} for details.}
    \label{fig:scattered}
\end{figure*}

After iterating through all the traces in all orders, the final RV for each band (red or blue) was determined as the median of the RVs from all traces, treating each trace within the same order as an independent measurement. As shown in the left panel of Figure~\ref{fig:scattered}, we plot the RVs derived from one observation for the blue and red bands as dark blue and red-filled circles, respectively (with peak SNR=600 and R=120,000, no contamination added). Each order has three individual traces (thus three RV points), and the final RV result in the two bands for this observation is the median of all the filled circles. We repeated this process for 100 observations, with the extracted RVs from all traces displayed as small light-colored points in the background. In the right panel of Figure~\ref{fig:scattered}, we present the derived RVs from these 100 observations for the blue and red bands. We then derived the RV precision for each band as the standard deviation of their respective distribution. In this example set of observations with SNR=600, R=120,000, and no contamination added, the RV is $6.21\mathrm{cm/s}$ in the blue band and $10.30\mathrm{cm/s}$ in the red band. The RV precision in the red band is slightly higher than in the blue band, driven by the greater Doppler information in the blue region of the solar spectrum (see Figure~\ref{fig:intro}).

\section{Results} \label{sec:res}

In this section, we present our results on quantifying the impact of fiber cross contamination on RV precision. We first show the RV precision in pure photon-limited scenarios as a validation of our pipeline. We then focus on the effect of the cal-sci contamination and sci-sci contamination separately. We combine all effects together in the last subsection. Again, we adopt the optical settings in the preliminary design of CHORUS as an example to illustrate these results.


\subsection{Photon-limited RV Precision \label{subsec:photonnoise}}

\begin{figure*}[ht]
    \centering
    \includegraphics[width=0.8\linewidth]{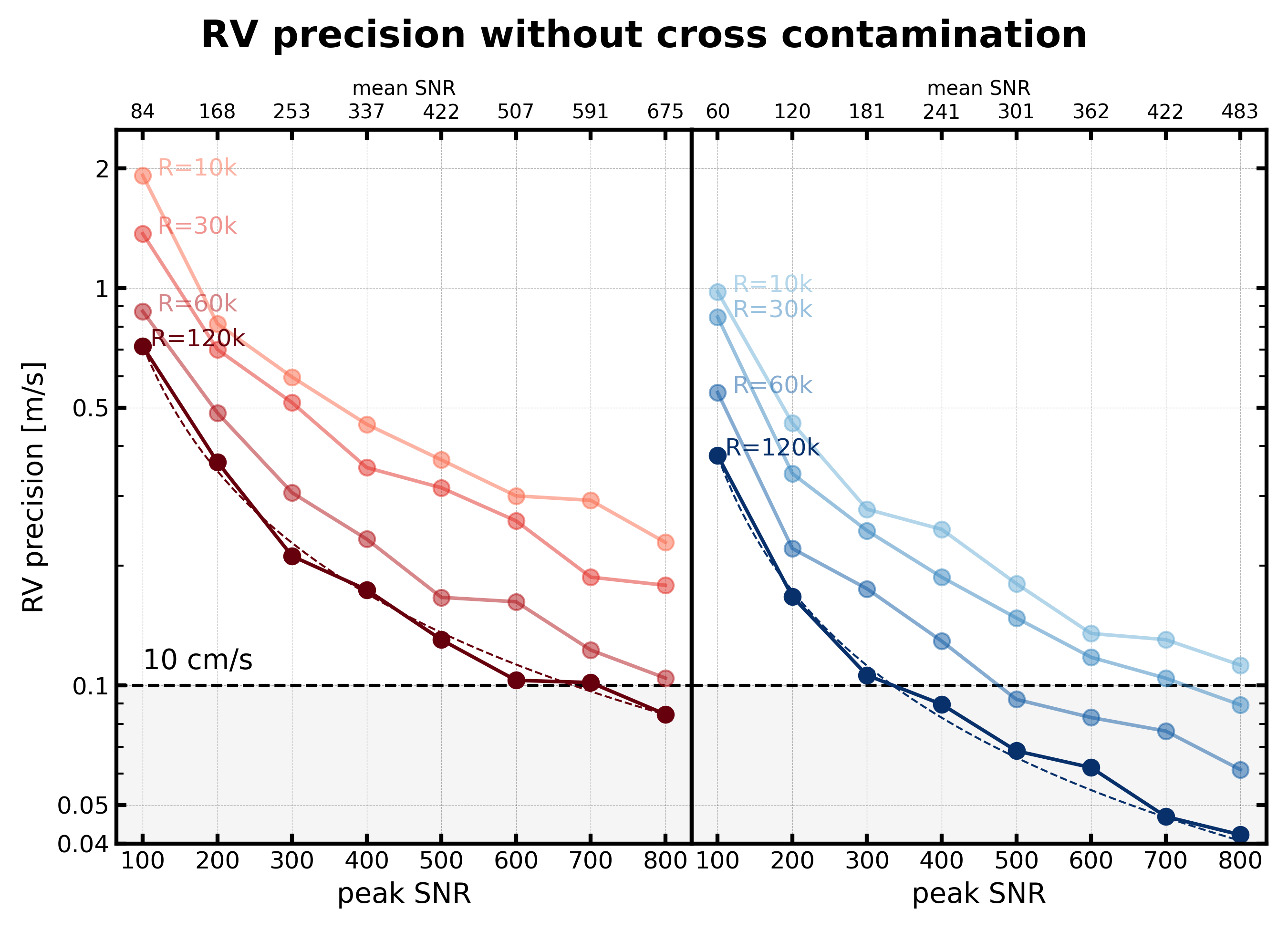}
    \caption{RV precision vs.\ peak SNR for different spectral resolutions, with the left plot showing the results for the red band and the right for the blue. The corresponding mean SNRs are labeled on the top x-axis. The dashed lines are $\sigma_{\text{RV}} \propto \text{SNR}^{-1}$. See Section~\ref{subsec:photonnoise} for details.}
    \label{fig:std-SNR}
\end{figure*}

\begin{figure*}[ht]
    \centering
    \includegraphics[width=0.8\linewidth]{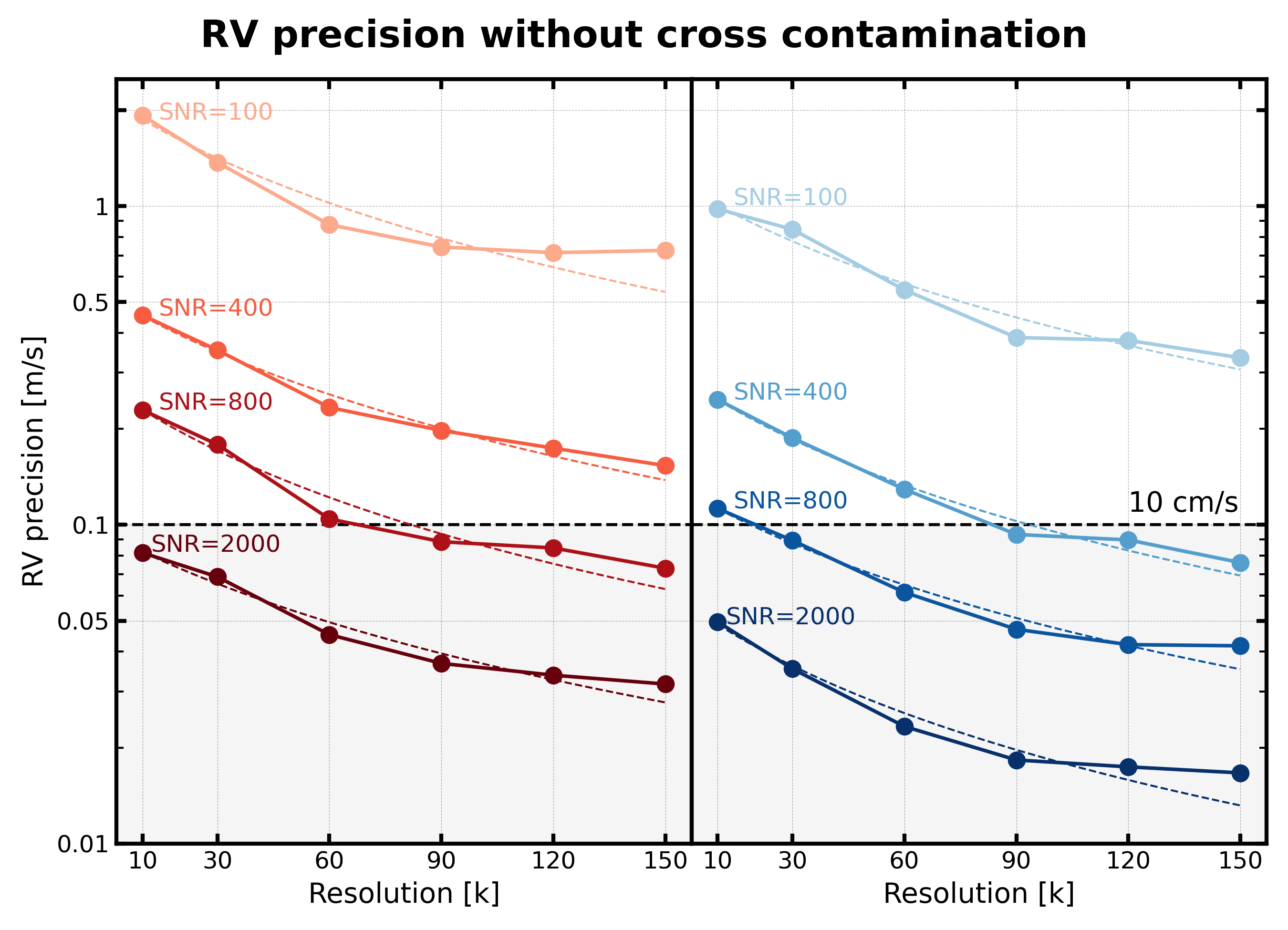}
    \caption{RV precision vs.\ spectral resolution for different SNRs, with the left plot showing results for the red band and the right for the blue. The dashed lines are the empirical scaling relation $\sigma_{\text{RV}} \propto R^{-1.21}$ \citep[e.g.,][]{2013PASP..125..240B}. Here, ``SNR"s in the plot are in short of ``peak SNR", the same as the x-axis in Figure~\ref{fig:std-SNR}. See Section~\ref{subsec:photonnoise} for details.}
    \label{fig:std-res}
\end{figure*}

We independently evaluated the effect of spectral resolution and SNR on the RV precision for three types of stars: a solar twin, a K dwarf with $T_\mathrm{eff}=4500\mathrm{K}$, and an M dwarf with $T_\mathrm{eff}=3700\mathrm{K}$, using the instrumental setups of CHORUS. We adopted six spectral resolution values --- 10,000, 30,000, 60,000, 90,000, 120,000, and 150,000. Additionally, we considered nine peak SNR values ranging from 100 to 800 in increments of 100, as well as an extreme value of 2000 to probe the systematic floor. We simulated 100 observations for each setting, and the results for the solar-type star are illustrated in Figure~\ref{fig:std-SNR} and Figure~\ref{fig:std-res}.

Figure~\ref{fig:std-SNR} shows the RV precision for the red and blue bands as a function of the SNR in the left and right panels, respectively. Results for different spectral resolutions are represented by lines of varying colors. The light gray region indicates an RV precision of below $10\mathrm{cm/s}$. Dashed lines show the results fitted with an empirical scaling relation, $\sigma_\mathrm{RV} \propto \mathrm{SNR}^{-1}$. As expected, RV precision improves (i.e., with decreasing standard deviations) with increasing SNR and resolution, consistent with the empirical scaling relation. To achieve a photon-limited RV precision of $10 \mathrm{cm/s}$ with spectral resolution R=120,000 (CHORUS preliminary design), we found that a peak SNR of approximately 400 (corresponding to a mean SNR of around 241) is required in the blue band, and a peak SNR of approximately 600 (with mean around 507) is needed in the red band. 

Figure~\ref{fig:std-res} presents the RV precision as a function of spectral resolution, again with separate panels for the red and blue bands. Results for different peak SNRs are represented by lines in various colors. As expected, our results follow the scaling relationship of $\sigma_{\text{RV}} \propto R^{-1.21}$ (e.g., \citealt{2013PASP..125..240B}). We also included results for an unrealistically high peak SNR of 2000 to probe the systematic floor of the RV precision (mostly set by the spectral resolution and the spectral sampling), which corresponds to mean SNR values exceeding 1200 in both bands that would saturate the CCD in a single exposure. As expected, the gain in RV precision slows down at higher SNRs. It also shows a hint of diminishing returns in precision gain when going beyond R=120,000, where the stellar spectral lines would all be well resolved (i.e., the solid lines being above the dashed lines beyond this resolution).

The results from the photon-limited simulations presented in this subsection will later be used to feed into the exposure time calculator of CHORUS to help estimate the expected RV precision and guide observational designs.

\subsection{Calibration-to-Science Contamination \label{subsec:calsci}}

We first quantified the influence on RV precision solely by cal-sci contamination. Throughout this section, we fixed the spectral resolution at R=120,000 and set the peak photon counts to 10,000 for the convenience of adding in the blaze, but we did not introduce any photon noise or readout noise so that we could isolate the effect of cal-sci contamination. We tested five values of cal-sci contamination fractions, logarithmically evenly spaced from 0.001\% to 10\%. Since no other random noise was included, we only conducted one simulated observation run for each setting. 

\begin{figure*}[ht]
    \centering
    \includegraphics[width=0.7\linewidth]{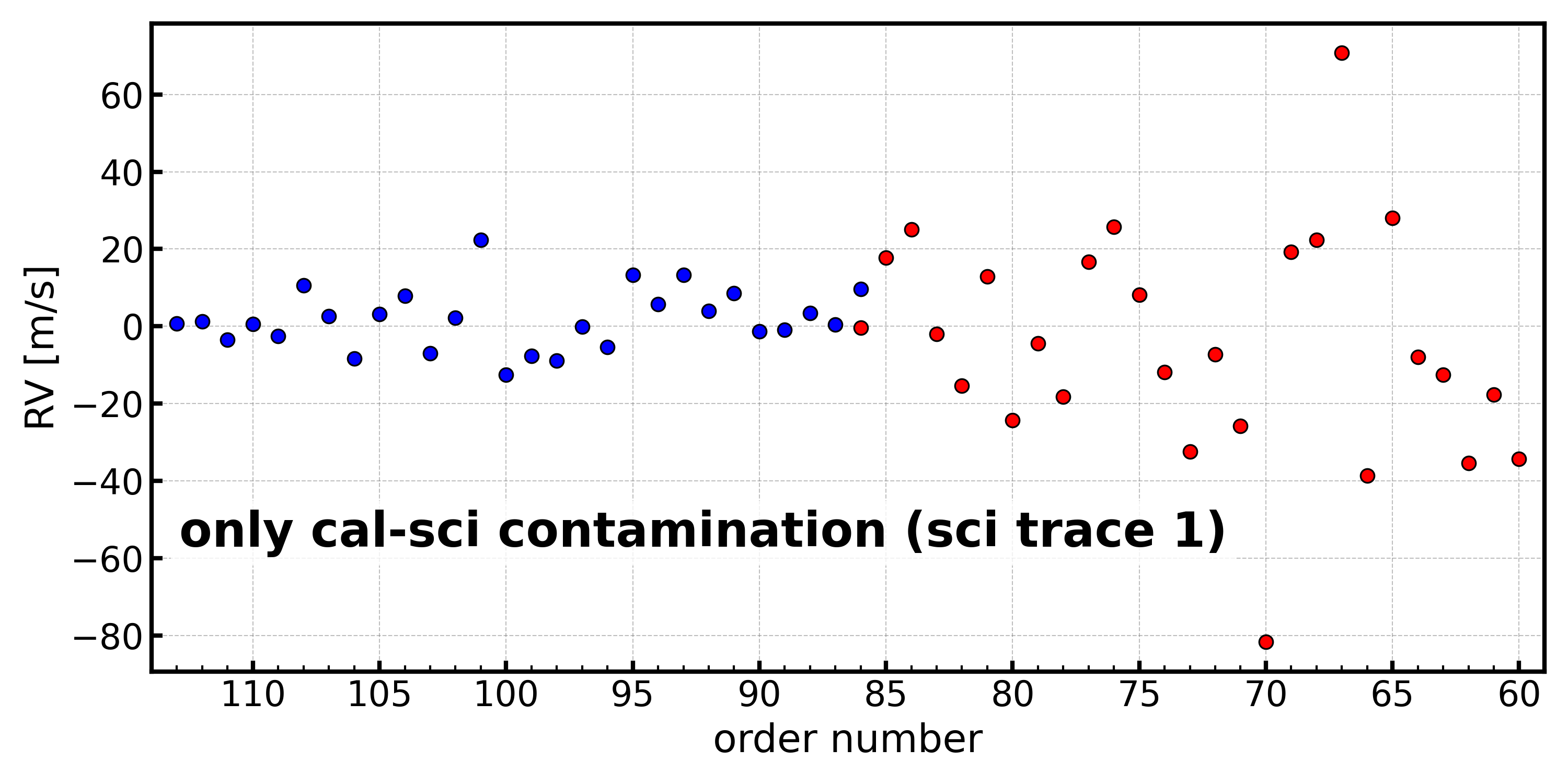}
    \caption{Scatter plot showing how the RVs are affected by cal-sci contamination. Each RV point is from the first trace in each spectral order, which is the trace closest to the calibration traces and thus contaminated. The spectrum was simulated with a spectral resolution of 120,000 and peak photon counts of 10,000, without adding any photon noise or sci-sci contamination. We added an unrealistically large contamination fraction of 10\% to amplify the effects for illustration purposes (same as in the top panel of Figure~\ref{fig:contamination_pixel}). See Section~\ref{subsec:calsci} for details.}
    \label{fig:scatter_calsci}
\end{figure*}

Figure~\ref{fig:scatter_calsci} illustrates the derived RVs of the influenced science trace 1 for both the blue and red bands under an unrealistically high contamination fraction of 10\% (the same as in the left plot of Figure~\ref{fig:contamination_pixel}). The results indicate that cal-sci contamination introduces both an RV offset (a bias) and additional RV scatter to the uncontaminated spectra. Moreover, the additional RV scatter is more pronounced in the red band compared to the blue band due to the differing levels of Doppler information in each band. We found a positive correlation between the absolute RV result of each trace and the ratio of the amount of Doppler information content between calibration and science spectra for that trace (calculated following Eqn.~6 in \citealt{1996PASP..108..500B}), suggesting that science traces with a higher density of spectral lines are more resilient against cal-sci contamination. From the extracted RVs of all orders, we derived the absolute RV offset as the median of all RVs in each band (median of the filled circles in Figure~\ref{fig:scatter_calsci}) and the induced RV scatter as the standard deviation.

\begin{figure}[ht]
    \centering
    \includegraphics[width=1\linewidth]{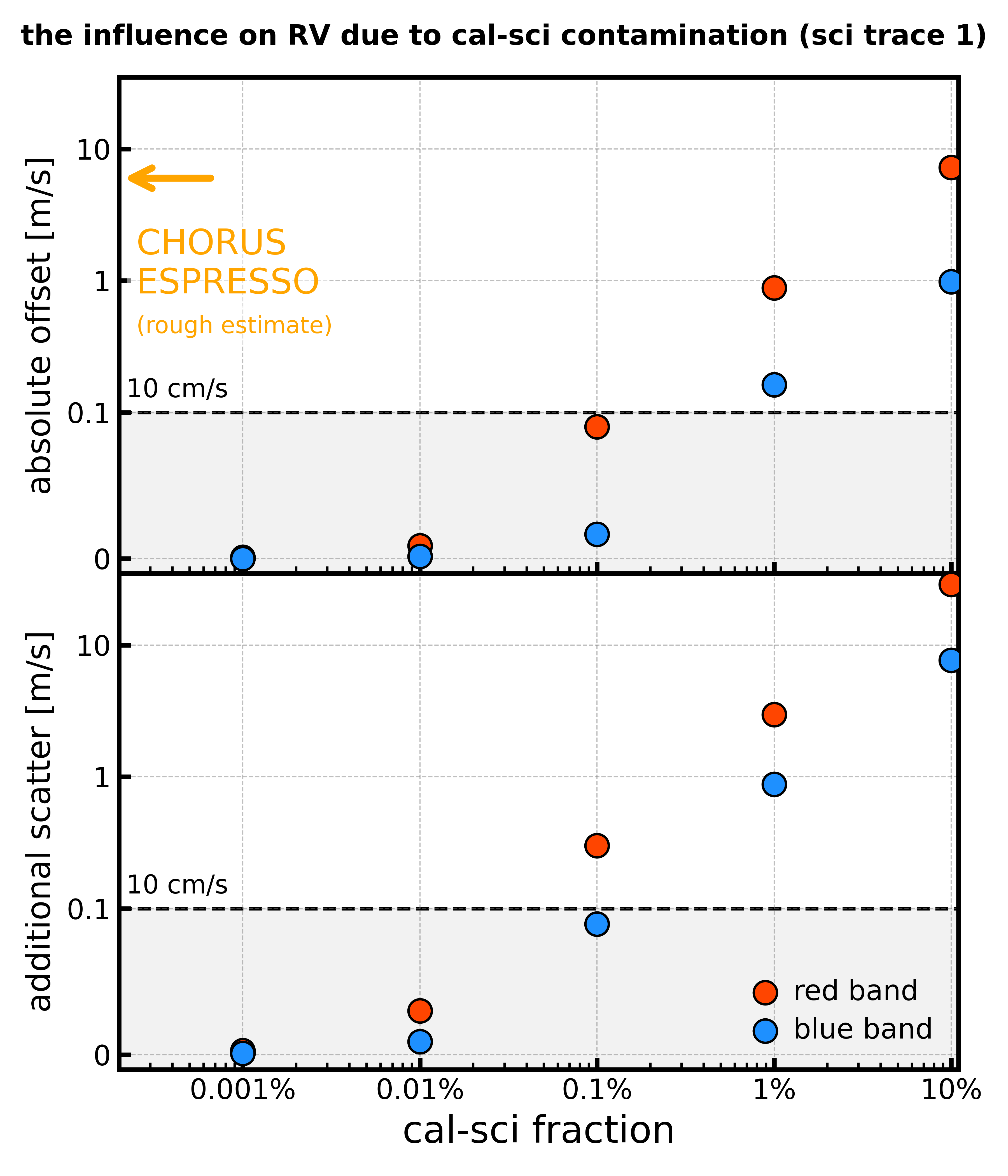}
    \caption{The absolute value of the RV offset and the additional scatter of science trace 1 introduced by cal-sci contamination for different amounts of contamination. Blue and red filled circles represent the results for the blue and red bands, respectively. Under a CHORUS or ESPRESSO-like setting, the typical cal-sci cross contamination fraction would be significantly smaller than 0.001\% and thus have a negligible impact on the final RVs. See Section~\ref{subsec:calsci} for details.}
    \label{fig:results_calsci}
\end{figure}

In Figure~\ref{fig:results_calsci}, we showed the influence of the cal-sci contamination fraction on the absolute RV offset and additional RV scatter in the upper and lower panels, respectively. The y-axis is presented on a logarithmic scale, with photon-limited RV precision of $<10 \mathrm{cm/s}$ highlighted in the light gray region. As expected, larger cal-sci contamination fractions lead to greater offsets and additional RV scatter. We also examined the order separation for both CHORUS preliminary design and ESPRESSO, and we found that the cal-sci contamination fraction for the narrowest separation\footnote{For CHORUS, the narrowest separation between two neighboring traces is 13 clear pixels with contamination fraction below 0.001\%, and the widest one can have more than 20 clear pixels.} was much lower than 0.0001\% — one order of magnitude lower than the lowest fraction we introduced (0.001\%, dashed grey line). We conclude that the cal-sci contamination is negligible under a CHORUS or ESPRESSO-like setting.

\subsection{Science-to-Science Contamination \label{subsec:scisci}}

We quantify the influence caused solely by sci-sci contamination in this subsection. For consistency, we fixed the spectral resolution R=120,000 and the peak photon counts at 10,000 while excluding photon random noise and readout noise. We selected a set of five sci-sci contamination fractions, logarithmically spaced from 0.001\% to 10\%, to evaluate the effects. Since random noise was excluded, we only conducted one simulated observation run for each setting. 

\begin{figure*}[ht]
    \centering
    \includegraphics[width=0.7\linewidth]{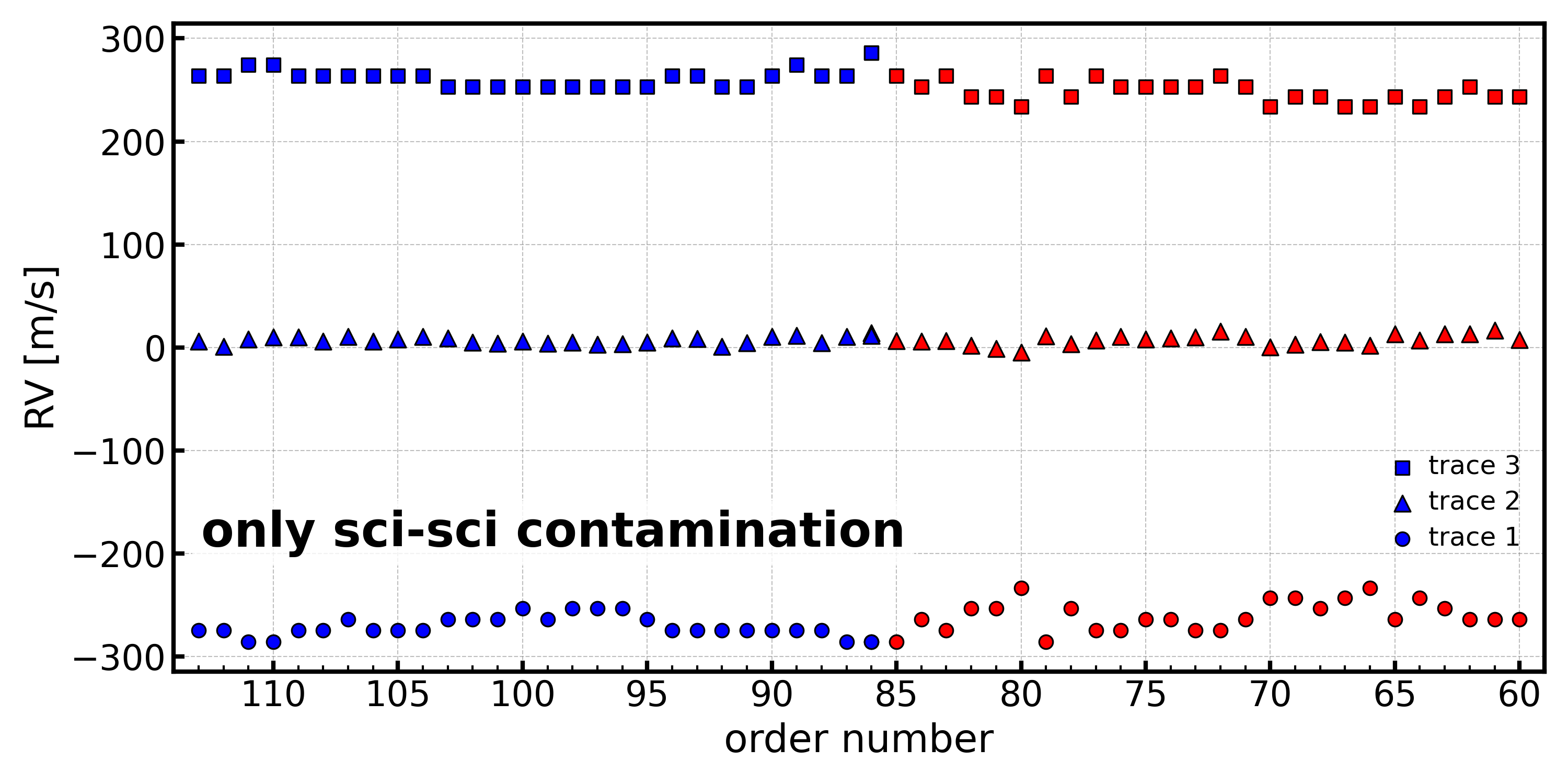}
    \caption{Scatter plot showing how the RVs are affected by sci-sci contamination. The three sets of RVs are for the three traces in each order in the blue and red bands in one simulated observation with sci-sci cross contamination fraction of 10\% (same as bottom panel of Figure~\ref{fig:contamination_pixel}). The circle, triangle, and square markers represent traces 1, 2, and 3, respectively (same as labeled in Figure~\ref{fig:ccd}). The spectrum was simulated with R=120,000 and peak photon counts of 10,000, without adding any photon noise. See Section~\ref{subsec:scisci} for details.}
    \label{fig:scatter_scisci}
\end{figure*}

In Figure~\ref{fig:scatter_scisci}, we plotted the derived RVs of three science traces under a 10\% sci-sci contamination fraction, represented by circle, triangle, and square markers for traces 1, 2, and 3, respectively. As shown in the lower panel of Figure~\ref{fig:contamination_pixel}, the sci-sci contamination to trace 1 from trace 2 distorts the line toward the left, causing the line center to shift slightly blueward. Consequently, the median RVs of trace 1 is negative (blue-shifted) in Figure~\ref{fig:scatter_scisci} (blue and red filled circles). Similarly, the sci-sci contamination to trace 3 from trace 2 distorts the line toward the right, causing the line center to shift slightly redward and the final median RVs to red-shifted (filled squares). For trace 2, however, the sci-sci contamination results from a combined effect of trace 1 distorting the line to the red and trace 3 distorting it to the blue, leading to a median RV around approximately zero (filled triangles). Unlike cal-sci contamination, there is no significant difference in additional RV scatter between the blue and red bands, since sci-sci contamination primarily affects the shape of all spectral lines in the same way, regardless of the intrinsic Doppler information within each order.

\begin{figure*}[ht]
    \centering
    \includegraphics[width=0.9\linewidth]{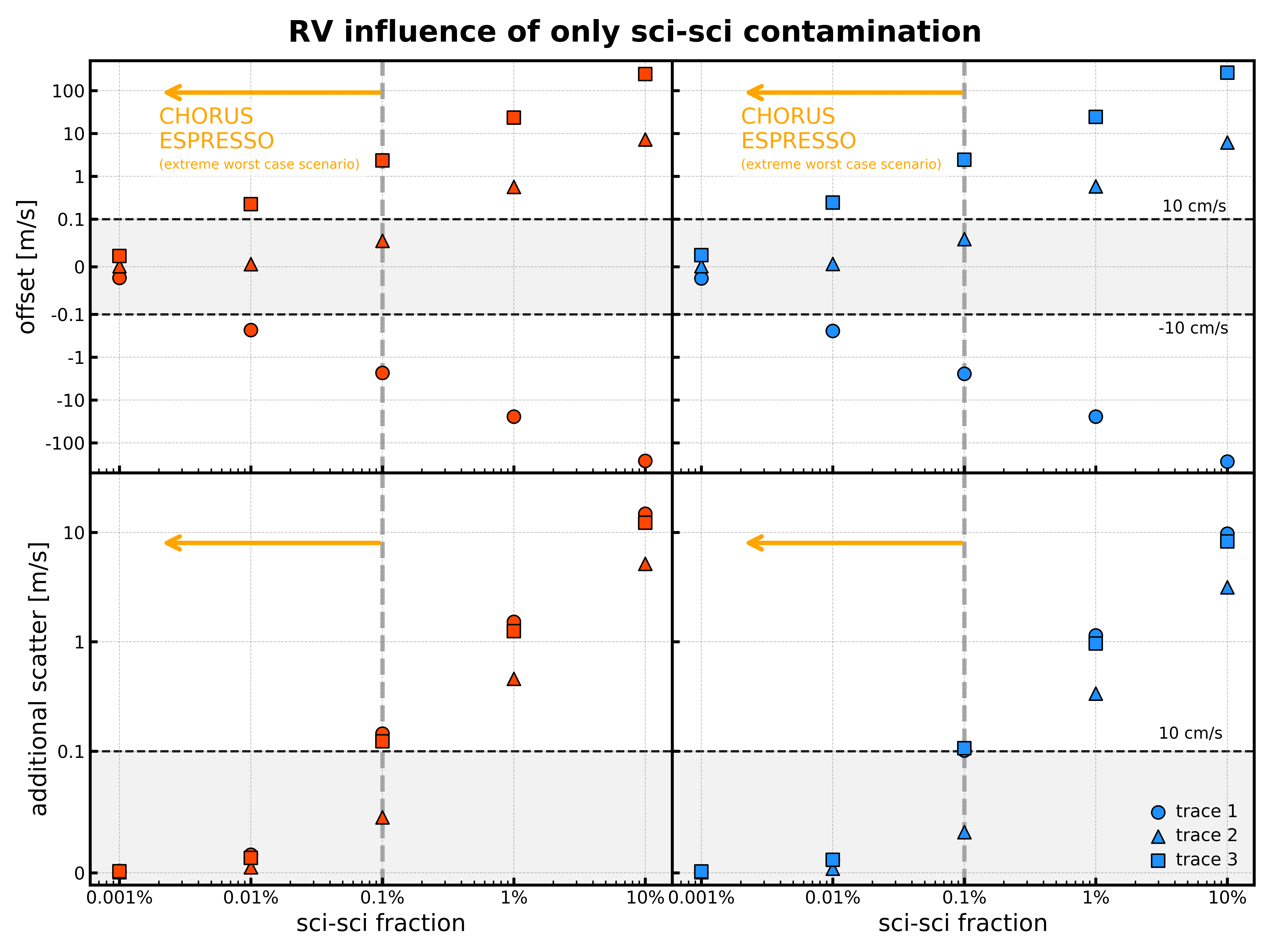}
    \caption{The RV offset and the additional RV scatter introduced by sci-sci contamination for different amounts of contamination. Blue and red filled markers represent the results for the blue and red bands, with circles, triangles, and squares indicating the results of science traces 1, 2, and 3 (same as Figure~\ref{fig:scatter_scisci}). The gray vertical line marked by the orange arrow illustrates an extreme worst-case scenario for the amount of sci-sci fraction for CHORUS and ESPRESSO, based on the separations between three science traces in each order. See Section~\ref{subsec:scisci} for details.}
    \label{fig:results_scisci}
\end{figure*}

In Figure~\ref{fig:results_scisci}, we present the influence of sci-sci contamination on the RV offset (shown as extracted RV values, rather than the absolute RV values, to highlight the distinct Doppler shifts) and additional scatter for the red and blue bands, respectively. As expected, larger sci-sci contamination fractions result in greater RV offsets and additional RV scatter. The offset for each trace in a single observation is trace-specific and can be mitigated by averaging the RVs of all three traces. We examined the trace separation for both CHORUS preliminary design and ESPRESSO, and we found that the sci-sci contamination fraction could be up to between 0.01\% and 0.1\% (corresponding to $y_\mathrm{c}$ ranging from 13 to 15 pixels based on CHORUS setting). This corresponds to an RV offset of up to $\sim$m/s level in some traces (though significantly smaller when averaged out among traces), as well as an additional scatter of up to $\sim$10~cm/s, which would contribute significantly to the RV error budget for PRV spectrographs like ESPRESSO and CHORUS.

However, we emphasize that these estimates are based on the assumption that \textit{all pixels} between traces were included in the spectral extraction window, which would overestimate the amount of cross contamination, so the fractions quoted here are upper limits. The grey dashed line and the leftward arrow in the figure indicate this estimated upper-limit fraction for CHORUS and ESPRESSO and the corresponding RV offset/scatter induced. In reality, there would be a couple to a few clean pixels between the extraction windows of neighboring traces, which would reduce the contamination fraction by at least an order of magnitude (see Figure~\ref{fig:cont_frac}), reducing the induced RV offset/scatter to a tolerable level, most likely. Background or scattered light subtraction would further reduce the cross contamination light in the extracted spectrum. In conclusion, careful design in fiber spacings and good practice in spectrum extraction are essential to minimizing sci-sci contamination in PRV spectra for instruments like CHORUS and ESPRESSO.

\subsection{Considering All Effects Together \label{subsec:chorus}}

\begin{figure*}[ht]
    \centering
    \includegraphics[width=0.7\linewidth]{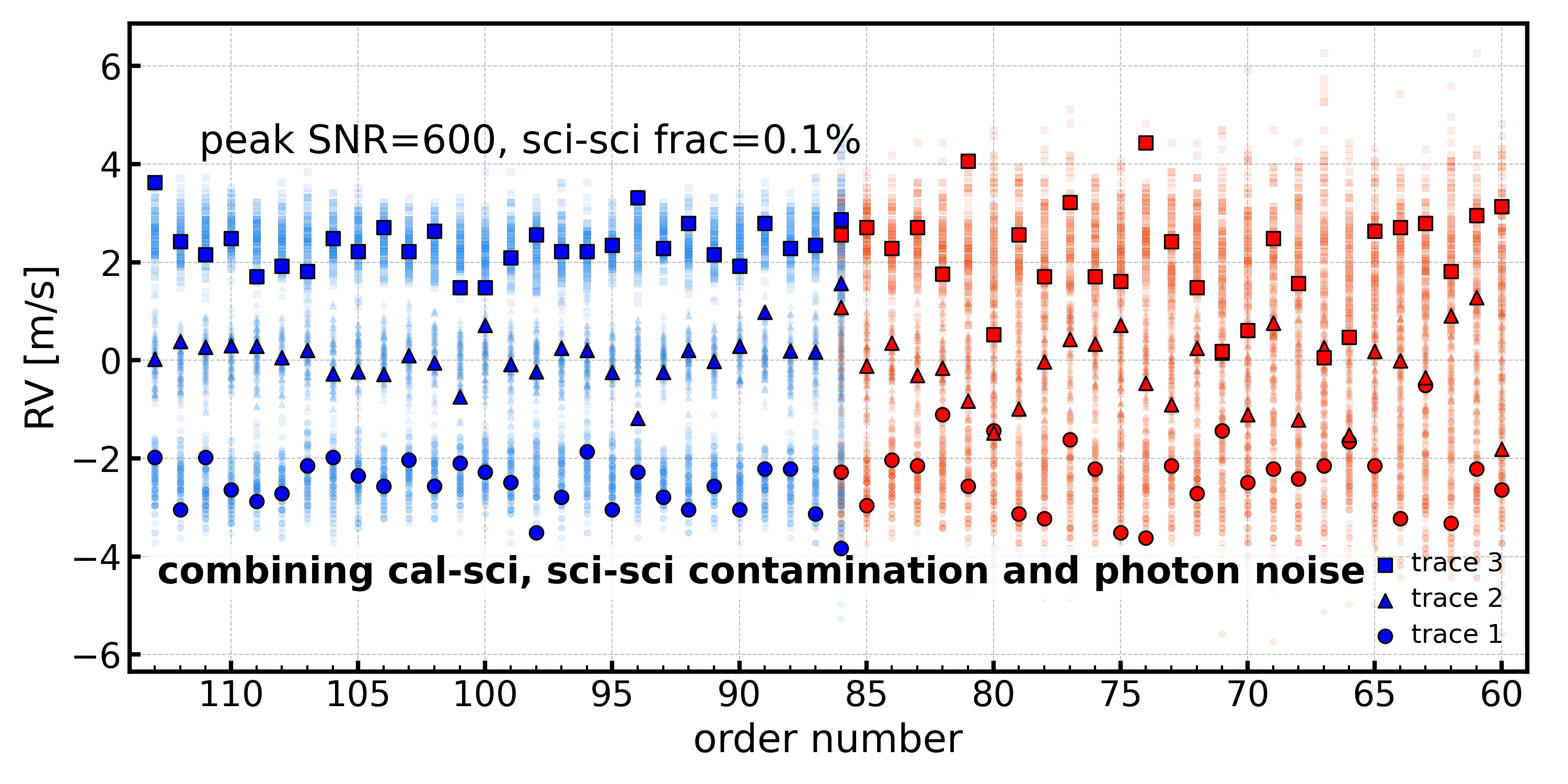}
    \caption{Scatter plot showing the RVs derived from each of the three traces in each spectral order throughout the 100 simulated observations (similar with Figure~\ref{fig:scattered}), taking into account photon noise, cal-sci contamination, and sci-sci contamination. The circle, triangle, and square markers represent traces 1, 2, and 3, respectively. The spectra were simulated based on the CHORUS preliminary design, with R=120,000, a peak SNR of 600, and some nominal assumptions on the amount of cross contamination (similar to Figures~\ref{fig:results_calsci} \& \ref{fig:results_scisci}) with a cal-sci contamination fraction of 0.0001\% and a sci-sci contamination fraction of 0.1\%. See Section~\ref{subsec:chorus} for details.}
    \label{fig:scatter_CHORUS}
\end{figure*}

\begin{figure*}[ht]
    \centering
    \includegraphics[width=0.8\linewidth]{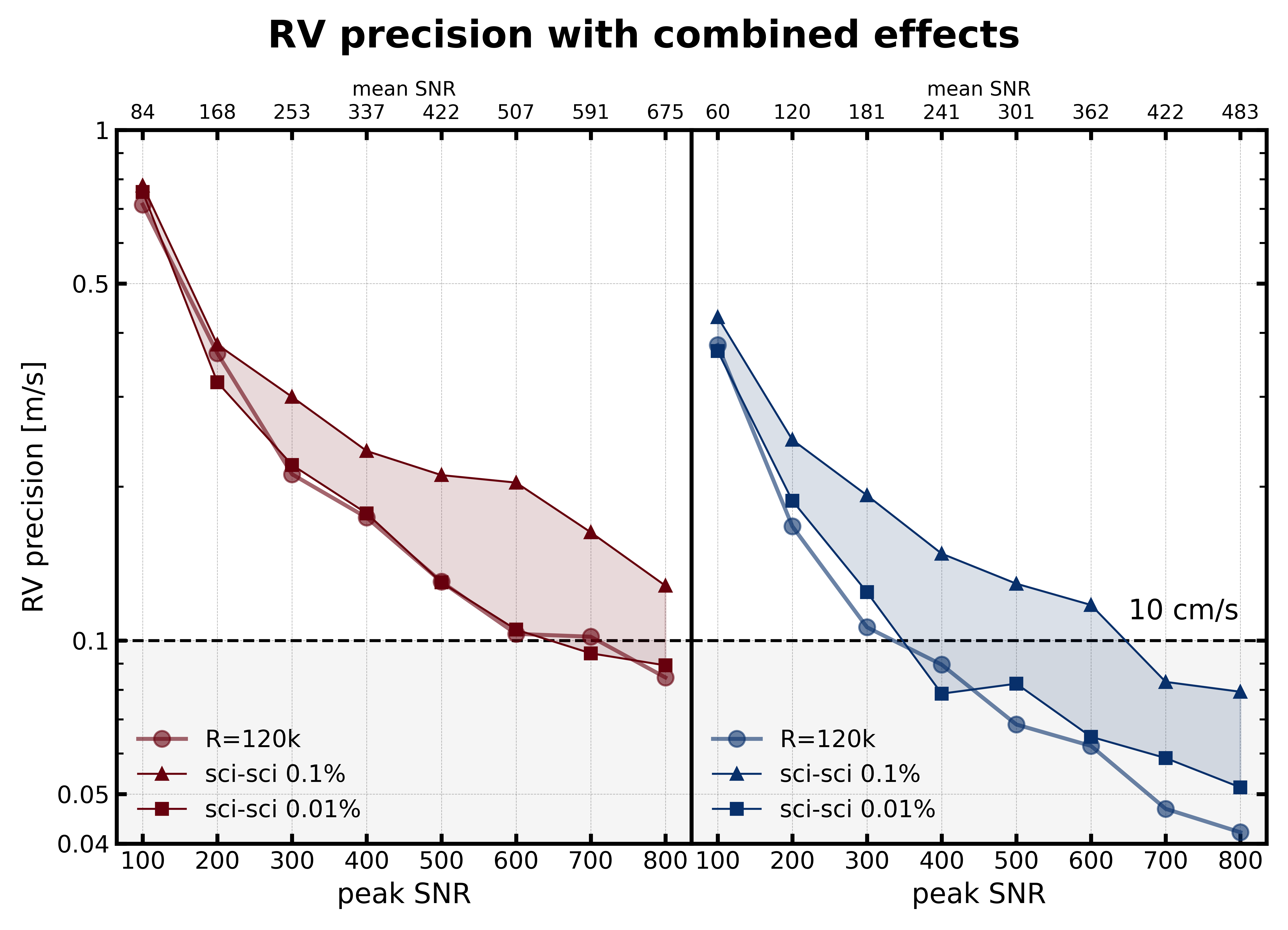}
    \caption{RV precision in the red/blue bands as a function of peak SNR for simulated time series of spectra with R=120,000, a fixed cal-sci contamination fraction of 0.0001\%, and varying sci-sci contamination fractions of 0.01\% and 0.1\%, based on the CHORUS preliminary design with some nominal assumptions for cross contamination (the same as Figure~\ref{fig:scatter_CHORUS}). The filled circles represent the results without contamination (corresponding to the darkest lines in Figure~\ref{fig:std-SNR}), while the filled triangles and squares represent the results at different sci-sci contamination fractions. See Section~\ref{subsec:chorus} for details.}
    \label{fig:results_CHORUS}
\end{figure*}

Finally, we estimate the RV precision under the combined influence of random noise (photon$+$readout) and the two types of contaminations. Again, we fixed the spectral resolution at R=120,000 and considered eight SNR values ranging from 100 to 800 in increments of 100. A small cal-sci contamination fraction of 0.0001\% was set, along with two sci-sci contamination fractions of 0.01\% and 0.1\% (see estimates for ESPRESSO and CHORUS preliminary design in the previous subsections; again, these are upper limits). We ran 100 simulated observations for each set of simulation parameters, and the results using the solar template are presented in Figure~\ref{fig:scatter_CHORUS} and Figure~\ref{fig:results_CHORUS} (see Section~\ref{subsec:comparison} for a comparison with K and M dwarfs).

In Figure~\ref{fig:scatter_CHORUS}, we plotted the RVs derived from the first observation for the blue and red bands using filled markers in dark blue and dark red, respectively, for a set of simulations with peak SNR of 600 and a sci-sci contamination fraction of 0.1\%. Circle, triangle, and square markers represent RVs from traces 1, 2, and 3, respectively. Light-colored points in the background represent RVs from the remaining 99 observations. The figure shows a clear non-zero RV offset between the traces due to sci-sci contamination with an amplitude $\sim 2$m/s, which exceeds the photon random noise at this relatively high SNR. The effect of cal-sci contamination is negligible in these sets of simulations given the very small contamination fraction (0.0001\%).

In Figure~\ref{fig:results_CHORUS}, the RV precision for the red and blue bands is shown as a function of peak SNR in the left and right panels, respectively. Different markers and shaded regions represent results for two different sci-sci contamination fractions (triangles and squares), while results without contamination are shown with transparent circles for comparison, corresponding to the bottom lines in Figure~\ref{fig:std-SNR}. We found that the results with a sci-sci contamination fraction of 0.01\% are consistent with those without contamination, indicating that a sci-sci contamination fraction below 0.01\% has a negligible impact on the final RVs. However, with all other settings remaining the same, a sci-sci contamination fraction of 0.1\% introduces additional RV scatter of approximately $\sim 10\mathrm{cm/s}$ at peak SNR around 400. Overall, these findings highlight the importance of spacing out the fiber traces and careful spectra extraction to mitigate the impact of cross contamination in RV spectroscopy.

\section{Discussion and Conclusion} \label{sec:dc}

\subsection{Comparison with Results Using K and M Dwarf Templates \label{subsec:comparison}}

The results presented above are all from simulations using a solar spectral template. We analyzed the simulated spectra and RVs for a K dwarf ($T_\mathrm{eff} = 4500$ K) and an M dwarf ($T_\mathrm{eff} = 3700$ K), to examine how different stellar types (or Doppler information) affect the photon-limited precision and the impact of contamination. As above, the simulations assumed no additional errors (details in Section~\ref{subsec:simulate}) except for the contamination and photon noise. 

\begin{figure*}[ht]
    \centering
    \includegraphics[width=0.8\linewidth]{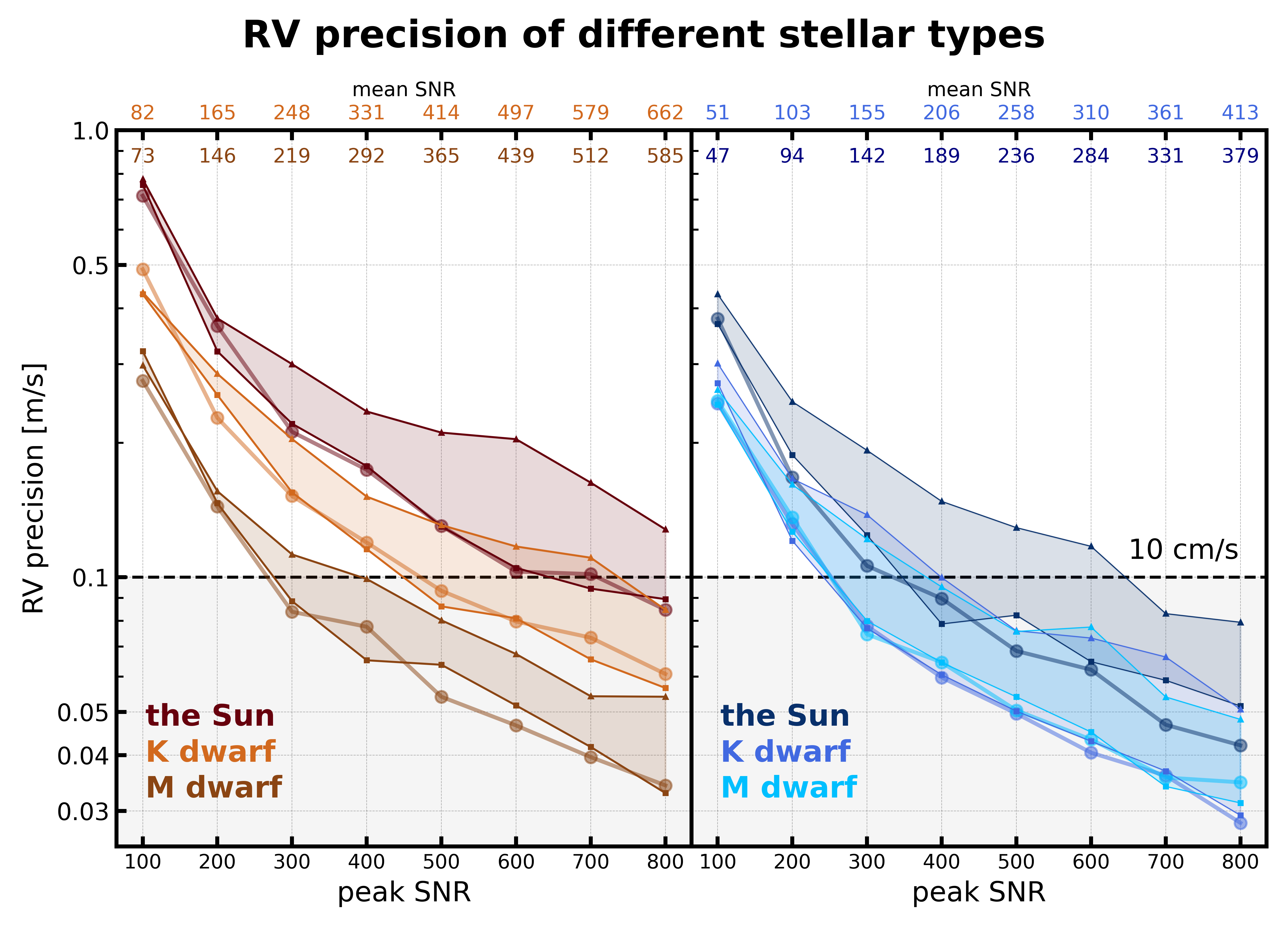}
    \caption{RV precision in the red and blue bands as a function of peak SNR for three input templates: the solar template, a K dwarf ($T_\mathrm{eff} = 4500$ K), and an M dwarf ($T_\mathrm{eff} = 3700$ K). Each template was annotated in different colors. The spectra were simulated as described in Section~\ref{subsec:input}. We annotated the mean SNRs, color-coded to match the lines, on the top x-axis (for the K dwarf at the top and the M dwarf below). See Section~\ref{subsec:comparison} for details.}
    \label{fig:discussion}
\end{figure*}

Figure~\ref{fig:discussion} shows the RV precision in the red and blue bands as a function of the peak SNR for different stellar types. Thicker transparent lines represent results without contamination (photon-limited), while shaded areas illustrate the impact of contamination, as described in Figure~\ref{fig:results_CHORUS} in Section~\ref{subsec:chorus}. In the red band, both K and M dwarfs have more spectral lines and thus greater Doppler information compared to the Sun, which results in better precision. However, in the blue band, the difference between K and M dwarfs is less pronounced. In conclusion, M and K dwarfs would suffer similarly as the G dwarfs from sci-sci contamination, although they offer higher RV precisions in general at the same SNR, thanks to a denser set of spectral lines, especially for the M dwarfs. However, our simulations did not account for real challenges in data reduction for M dwarfs, such as potential additional errors in continuum normalization and stronger stellar activities, in general.

\subsection{Caveats \label{subsec:cf}}

In this section, we outline the caveats of our simulations regarding the impact of cross contamination and suggest directions for future work.

First, as discussed in Section~\ref{subsec:cont}, the separation between traces and orders varies across different positions on the CCD, causing the contamination ratio to fluctuate across pixels within an order and also between orders. This variation affects both cal-sci and sci-sci contamination, with a more pronounced impact on sci-sci contamination. In our simulations, we assumed a fixed, large contamination fraction to represent an extreme worst-case scenario for RV precision degradation due to sci-sci contamination. While this approach provides a rough illustration of the effects of cross contamination on the two instruments we chose (i.e., CHORUS and ESPRESSO), future refinements could incorporate a variable contamination fraction that changes with pixel position for more accurate modeling.

Second, we simplified the simulation by assuming (1) identical SNRs for calibration (LFC or FP) and science spectra, \revise{and (2) the same SNR for the three traces in each order. For assumption (1), i}n practice, the brightness of calibration lines can be adjusted according to different needs, and in cases where the science spectra have very low SNRs, the calibration lamps may even be turned off. However, since we focused on the regime with photon-limited RV precision close to the instrument's systematic floor, which typically involves science spectra with relatively high SNRs, we assumed that the calibration lamps would always be as bright as the science spectra. This simplification reduces complexity while maintaining relevance for the intended precision estimates. In reality, unnecessarily bright calibration light (or smaller separation between cal and sci fibers than assumed here) would cause more cal-sci contamination than estimated in this work, so careful practice in adjusting and monitoring the calibration light brightness is never a bad idea.

\revise{For assumption (2), we assumed identical SNRs for three traces in one order (or, to say, the pupil slicer evently divides the photons into three parts at the pupil) in Section~\ref{subsec:simulate}. Effectively, pupil slicers cannot strictly trisect the injected beam, and the sliced pupil is trisected along the major axis of this elliptical aperture. For example, for the preliminary design of CHORUS, the photon count ratio of three traces within one order is roughly $0.7:1:0.7$ (under idealized cases). Trace 2 is expected to hold the most photons (or have the highest SNR), and is the least influenced by cross contamination as well (shown in Figure~\ref{fig:results_scisci}). Therefore, our simulation might have slightly elevated the influence of trace 1 and trace 3 on RV precision, making the reported RV error an overestimate. Addressing this small variation in photon counts between traces in subsequent research would lead to more accurate results.}

\revise{Furthermore, we assumed a constant spectral resolution value of $R=120,000$ for one setting, while in reality, the resolution and PSF of CHORUS change with wavelength both within one order and across orders according to Eq.~\ref{eqn:sigma}. Resolution at the middle wavelength of all orders is around $120,000$. Within an order, spectral resolution increases from shorter to longer wavelengths, often with a considerable and measurable amount. Meanwhile, there is also an overall modest enhancement in resolution from the blue to the red band across all orders. We adopted a fixed value of $120,000$ for the spectral resolution as this is the typical resolution near the blaze peak of each order, representative for the resolution of the entire instrument. A more accurate simulation and estimate of the RV precision would require a changing resolution (and PSF) consistent with specific instrumental design within and across orders.}

Lastly, we assumed the spectral data reduction process to be simply a basic window extraction process with a very wide extraction width. Consequently, the RV precision results influenced by sci-sci contamination, as discussed in Section~\ref{subsec:scisci}, should be interpreted as an extreme worst-case scenario. Future works involving simulating the 2-D spectra on the CCD and the full spectral extraction process would make the results more accurate.

To facilitate further research, we have made our simulation code publicly available on \texttt{GitHub}\footnote{Code available at: \url{https://github.com/JoanneJi/2025-FiberCrossContamination/tree/master}.}. The code, designed with compatibility for common \texttt{Python} packages, can be adapted to refine the contamination model, support the exposure-time calculator for CHORUS, and provide rough estimates of photon-limited RV precision for other spectrographs. \revise{Additionally, users can change the input parameters of the code to account for a changing resolution across orders and uneven SNRs between traces.}



\subsection{Summary and Future Work \label{subsec:sc}}

In this work, we investigated the impact of fiber cross contamination on radial velocity precision, using the preliminary optical design of CHORUS as a test case. We simulated the extracted 1-D spectra under four scenarios: photon noise only (as a baseline comparison), cal-sci contamination only, sci-sci contamination only, and the combined effects of both contaminations and photon noise. Our findings highlight the importance of considering the fiber trace spacing and the fiber cross contamination, especially the sci-sci contamination, in the instrument designs and the spectral extraction process. We found that cal-sci contamination with a fraction of $0.1\%$ can lead to absolute RV offsets of up to a few cm/s and an additional RV scatter of up to $\sim 10\mathrm{cm/s}$. The current settings of both CHORUS and ESPRESSO are safe against cal-sci contamination thanks to a large separation between calibration and science fiber. In the worst-case scenario, with a large spectral extraction window, sci-sci contamination at $0.1\%$ can introduce additional blueshift or redshift in different traces, with a maximum amplitude of several meters per second and contributing up to $\sim$10~cm/s to the RV error budget, although these numbers could be an order of magnitude smaller with a more careful spectral extraction in practice. 

In general, for spectrographs with fiber spacings similar to the preliminary design of CHORUS (and also ESPRESSO), assuming a careful spectral extraction, fiber cross contamination should not contribute significantly to the RV error budget beyond $\sim$cm/s level. Since parallel calibration fibers and the adoption of a pupil or image slicer are common in modern high-precision RV spectrographs, we emphasize the importance of evaluating the effects of fiber trace separation and spectral extraction strategies to minimize contamination and maximize instrumental precision for current and future PRV spectrographs.

\vspace{0.8cm}
C.J. and S.X.W acknowledge the support from the Scientific Instrument Developing Project of the Chinese Academy of Sciences (Grant No. ZDKYYQ20220009), the international partnership program of the Chinese Academy of Science (Grant No. 178GJHZ2022047GC), and the Tsinghua Dushi Fund (53121000123, 53121000124). We thank the help from the CHORUS science and instrument team. C.J. thanks Zhecheng Hu and Zitao Lin for their valuable support for the spectra normalization process. C.J. also thanks Jiayue Zhang for her suggestions on making readable repositories on \texttt{GitHub}.This research has made use of the Astrophysics Data System (\url{https://ui.adsabs.harvard.edu/}), funded by NASA under Cooperative Agreement 80NSSC21M00561.  





\software{\texttt{Astropy} \citep{astropy:2013, astropy:2018, astropy:2022},
        \texttt{NumPy} \citep{harris2020array},
        \texttt{Matplotlib} \citep{Hunter:2007},
        \texttt{iSpec} \citep{2014A&A...569A.111B, 2019MNRAS.486.2075B},
        \texttt{SciPy} \citep{2020SciPy-NMeth},
        \texttt{Zemax}
          }

\newpage
\bibliography{PASPsample631}{}
\bibliographystyle{aasjournal}



\end{document}